\documentclass[a4paper,11pt]{article}

\usepackage{jheppub} 

\usepackage[T1]{fontenc} 

\usepackage{graphicx}
\usepackage{tikz}
\usepackage{braket}

\input{arrowsnew.tex}

\def\a{\alpha}
\def\b{\beta}
\def\d{\delta}

\def\e{\epsilon}

\def\k{\kappa}
\def\l{\lambda}

\def\m{\mu}

\def\s{\sigma}

\def\ph{\phi}

\def\th{\theta}

\def\ch{\chi}
\def\abs#1{\left| #1 \right|}
\def\pd{\partial}
\def\avg#1{\langle #1 \rangle}

\def\br{\mathbf{r}}
\def\bu{\mathbf{u}}
\def\bau{\bar{u}}

\def\ui{\{u_i\}}
\def\hnui{H_N(t,\{u_i\})}

\def\bfx{\mathbf{x}}

\def\scM{\mathcal{M}}

\def\scO{\mathcal{O}}

\newcommand{\bigslant}[2]{{\left.\raisebox{.2em}{$#1$}\middle/\raisebox{-.2em}{$#2$}\right.}}

\newcommand{\be}{\begin{eqnarray}}
\newcommand{\ee}{\end{eqnarray}}
\newcommand{\dd}{{\partial}}

\makeatletter
\newsavebox\myboxA
\newsavebox\myboxB
\newlength\mylenA
\newcommand*\xoverline[2][0.75]{%
    \sbox{\myboxA}{$\m@th#2$}%
    \setbox\myboxB\null
    \ht\myboxB=\ht\myboxA%
    \dp\myboxB=\dp\myboxA%
    \wd\myboxB=#1\wd\myboxA
    \sbox\myboxB{$\m@th\overline{\copy\myboxB}$}
    \setlength\mylenA{\the\wd\myboxA}
    \addtolength\mylenA{-\the\wd\myboxB}%
    \ifdim\wd\myboxB<\wd\myboxA%
       \rlap{\hskip 0.5\mylenA\usebox\myboxB}{\usebox\myboxA}%
    \else
        \hskip -0.5\mylenA\rlap{\usebox\myboxA}{\hskip 0.5\mylenA\usebox\myboxB}%
    \fi}
\makeatother

\def\bou{\xoverline{u}}

\usetikzlibrary{positioning,decorations.pathreplacing,decorations.markings,shapes}
\usetikzlibrary{calc}
\usetikzlibrary{arrows,shapes,backgrounds}
\usetikzlibrary{decorations.pathmorphing}
\usetikzlibrary{decorations.markings}

\tikzset{->-/.style={decoration={
  markings,
  mark=at position #1 with {\arrow{latex'new}}},postaction={decorate}}}
\tikzset{->>-/.style={decoration={
  markings,
  mark=at position #1 with {\arrow{>}}},postaction={decorate}}}

\title{Hilbert series and operator bases with derivatives in effective field theories}

\author[a,b]{Brian Henning,}
\author[a,b]{Xiaochuan Lu,}
\author[a,b]{Tom Melia}
\author[a,b,c]{and Hitoshi Murayama}

\affiliation[a]{Department of Physics, University of California, Berkeley, California 94720, USA}
\affiliation[b]{Theoretical Physics Group, Lawrence Berkeley National Laboratory, Berkeley, California 94720, USA}
\affiliation[c]{Kavli Institute for the Physics and Mathematics of the Universe (WPI), Todai Institutes for Advanced Study, University of Tokyo, Kashiwa 277-8583, Japan}

\emailAdd{bhenning@berkeley.edu}
\emailAdd{luxiaochuan123456@berkeley.edu}
\emailAdd{tmelia@lbl.gov}
\emailAdd{hitoshi@berkeley.edu, hitoshi.murayama@ipmu.jp}

\preprint{UCB-PTH 15/05}
\preprinta{IPMU15-0112}

\abstract{We introduce a systematic framework for counting and finding independent operators in effective field theories, taking into account the redundancies associated with use of the classical equations of motion and integration by parts.  By working in momentum space, we show that the enumeration problem can be mapped onto that of understanding a polynomial ring in the field momenta. All-order information about the number of independent operators in an effective field theory is encoded in a geometrical object of the ring known as the Hilbert series. We obtain the Hilbert series for the theory of $N$ real scalar fields in $(0+1)$ dimensions---an example, free of space-time and internal symmetries, where aspects of our framework are most transparent. Although this is as simple a theory involving derivatives as one could imagine, it provides fruitful lessons to be carried into studies of more complicated theories: we find surprising and rich structure from an interplay between integration by parts and equations of motion and a connection with $SL(2,{\mathbb C})$  representation theory which controls the structure of the operator basis.}

\begin{document} 
\maketitle
\flushbottom

\section{Introduction}
The Wilsonian picture of effective field theory (EFT) dictates that the effective action contains all possible local operators consistent with the symmetries of the EFT,
\begin{equation}
S_{\text{eff}} = \int_{\scM} \sum_{a}\k_a \scO_a,
\label{eqn:Seff}
\end{equation}
where $\scO_a$ are local operators,  \(\k_a\) are their associated Wilson coefficients, and \(\scM\) is the spacetime manifold. It is well known that distinct operators can lead to the same physical effect, \emph{i.e.} give the same \(S\)-matrix elements. Two examples, which will play a prominent role in this work, are operators related by integration by parts (IBP) and operators related through the equations of motion (EOM). We will refer to the minimum set of operators encapsulating all possible physical effects as a \emph{basis} for the EFT.

The issue of determining (subsets of) the EFT basis has arisen in the past for various specific EFTs, but no general prescription exists. Logically, properties of the basis should be discussed first since they include, for example, the number of physically distinct effects. Determining the EFT basis is a very difficult task. Ideally, one wants some sort of generating function, if it exists, which encodes information about the generators of a basis together with all possible relations among these generators. In addition to allowing us to enumerate the operators in an EFT basis, such a function is appealing since it would encode information about the basis as a whole---it could potentially reveal properties of the EFT that are difficult or impossible to see when working with any fixed subset of operators.

To obtain a function with many of the aforementioned properties, we define a \emph{generalized Hilbert series} as follows. Associate a weight \(u_i\) to each field \(\ph_i\) in the EFT and a weight \(t\) to the (covariant) derivative. The generalized Hilbert series is then defined as
\begin{equation}
H(t,\{u_i\}) = \sum_{k=0}^{\infty}\sum_{r_1=0}^{\infty}\dots\sum_{r_N=0}^{\infty} c_{k\, r_1 \dots r_N} t^ku_1^{r_1}\dots u_N^{r_N},
\label{eqn:Hdef}
\end{equation}
where \(c_{k\, r_1 \dots r_N} \in {\mathbb N}\) is the number of independent operators in the basis composed of \(k\) derivatives and  \(r_1,\dots,r_N\) powers of the \(\ph_1, \dots, \ph_N\). The weights \(\{t,u_i\}\) are complex numbers; formally, in order to have the above series converge, we require \(\abs{t},\abs{u_i}< 1\).

The summed Hilbert series succinctly encodes information about the EFT basis. For simplicity, let us assume all fields in the EFT are bosonic. Denoting the (possibly infinite) set of generators of the EFT basis by \(X_{\text{gen}}\), the denominator of the Hilbert series, \(H = N/D\), takes the form
\begin{equation}
D(t,\{u_i\}) = \prod_a (1-h_a),
\end{equation}
where \(h_a\) is the weight of an operator \(\scO_a \in X_{\text{gen}}\). In the case where any operator in the basis can be \emph{uniquely} written as a polynomial of the generators we say the basis is \emph{freely generated}, \emph{i.e.} there are no non-trivial relations among the generators. In this case the numerator is unity: binomial expansion of the denominator clearly shows that the weights of all operators in the basis are obtained and counted once. A numerator of \(H\) differing from unity accounts for relations among the generators.

The grading where each field has a unique weight is, of course, just a choice; while this grading retains the most information about the basis, different weighting choices may be more expedient for other purposes. For example, we could grade by dimension of operators: let \(d_i = [\ph_i]\) be the dimension of \(\ph_i\) and send \(u_i \to q^{d_i}\) and \(t \to q\) so that the coefficient \(c_k\) in expansion of the Hilbert series, \(H(t,\{u_i\}) \to H(q) = \sum c_k q^k\), is the number of dimension \(k\) operators in the basis. 
In the case of a conformal theory, the operator-state correspondence implies that the Hilbert series when only including EOM coincides with the partition function when we grade operators by their scaling dimension, angular momentum, and other quantum numbers.\footnote{IBP removes local operators which are total derivatives, but these must be included in the partition function.}

Hilbert series are common objects in algebraic geometry and commutative algebra. They have appeared in the physics literature through their connection to
invariant theory,\footnote{A brief and accessible introduction to invariant theory and Hilbert series can be found in section IV of~\cite{Jenkins:2009dy}.}
finding application in, \emph{e.g.}, studying flavor invariants~\cite{Jenkins:2009dy,Hanany:2010vu} and counting gauge invariants in SUSY gauge theories~\cite{Benvenuti:2006qr,Feng:2007ur,Gray:2008yu}. 
A recent work reviewed and emphasised their utility for counting operators that do not contain derivatives in phenomenological
settings, such as in subsets of the Standard Model EFT~\cite{Lehman:2015via}. The present work can be seen as a contribution towards extending the use of Hilbert series in
 general EFTs by including operators with derivatives.
Explicitly, we show how to systematically deal with IBPs and EOM, both of which define equivalence relations:
\begin{enumerate}
\item \textbf{IBPs:} Two operators are equivalent if they are related by a total derivative,
\begin{equation}
\scO_a \sim \scO_b \quad \text{if} \quad \scO_a = \scO_b + \text{d} \scO_c,
\label{eqn:equiv_IBP}
\end{equation}
since \(\int_{\scM} \text{d} \scO = 0\).\footnote{As usual, the operators are assumed to vanish at infinity in the case where \(\scM\) is non-compact.}
\item \textbf{EOM:}
Two operators related by the EOM lead to the same physical effects~\cite{Politzer:1980me,Georgi:1991ch}. The EOM follow from the requirement
\be
\frac{\delta S_{\text{eff}}}{\delta \phi_i}=0 \,.
\ee
The above equation is often evaluated at lowest order in an expansion parameter of the EFT; the operator relations implied
are then worked out at each successive order.
For the purpose of simply counting operators, considering equations of motion derived from kinetic terms is sufficient. We therefore have the following equivalence,
\begin{equation}
\scO_a \sim \scO_b \quad \text{if} \quad \scO_a = \scO_b + \scO_c \frac{\d S_{\text{kin}}}{\d \ph_j},
\label{eqn:equiv_EOM}
\end{equation}
for some \(\ph_j \in \{\ph_i\}\).
\end{enumerate}

The physical system to which we restrict ourselves in this paper is a one-dimensional quantum field theory of \(N\) real scalar fields, \(\ph_1, \dots, \ph_N\). We refer to the \(N\) fields as flavors and denote the generalized  Hilbert series as \(H_N(t,\{u_i\})\). Even though this is the simplest theory one could imagine for our purpose, we find surprising and non-trivial results when both IBP and EOM are accounted for.

The outline of this paper is as follows.
In section~\ref{sec:framework} we develop a framework for computing the coefficients $c_{k\, r_1 \dots r_N}$ of the Hilbert series and constructing sets of independent operators, while taking into account IBP and EOM. Although this section is limited to one dimension only, the framework is general: it outlines in principle a systematic procedure for obtaining both the dimensionality and explicit realizations of an operator basis for an EFT. 

Sections~\ref{sec:intuitive_calc} and~\ref{sec:su2} explore the possibility of summing
the series and obtaining an all-orders result for the EFT considered here.
Section~\ref{sec:intuitive_calc} relies heavily on the formalism developed above, and we show that while the summed
series is easily obtainable when accounting for only IBP or EOM, taken together the problem becomes considerably complicated. We present
a sum formula for the Hilbert series for general $N$, for which we are able to perform the summation for relatively low values of $N$.
Section~\ref{sec:su2} exploits a connection with the representation theory of $SL(2,{\mathbb C})$ that is present in this theory. Using this formalism, we are able to obtain the closed form of the Hilbert series for general $N$. 

We go on to explore  the analytic structure of the Hilbert series in section~\ref{sec:poles}. This reveals interesting consistency conditions and recursion relations connecting different Hilbert series. We conclude in section~\ref{sec:summary} with a discussion aimed at highlighting the physical aspects of this work and the nature of its generalizations to EFTs in higher space-time dimensions. 

We include a basic introduction to  commutative algebra in appendix~\ref{app:a} and a demonstration implementing our framework in the computer package {\tt Macaulay2} in appendix~\ref{app:b}.


\section{Framework} \label{sec:framework}

In this section we provide a systematic framework for computing the Hilbert series, \emph{i.e.} counting operators modulo integration by parts and use of the equations of motion. Throughout this section, we restrict the discussion to our one-dimensional QFT with \(N\) flavors of real scalar fields $\phi_i$, \(i=1,\dots,N\). We take the spacetime manifold to be the circle \(S^1\) with coordinate \(\th\); the counting is the same here as for the real line. 

We take the equations of motion from \(\int_{S^1}d\th \sum_i (\pd\ph_i)^2/2\), which gives \(\pd^2\ph_i=0\). In one-dimension, terms of the form \(\ph_j\pd \ph_i\), \(i\ne j\), are present in the action and could be included in the EOM. However, they do not change the counting because \(\pd^2\ph_i\) is always in the EOM as long as \(\phi_i\) has a non-zero kinetic term.

Introducing a shorthand notation \(\bu^{\br}\) for  \(u_1^{r_1}\dots u_N^{r_N}\) and \(c_{k\, \br} = c_{k\, r_1\dots r_N}\), the Hilbert series is written as
\begin{equation}
H_N(t,\ui) = \sum_{k,\mathbf{r}=0}^{\infty}c_{k\, \mathbf{r}}\mathbf{u}^{\mathbf{r}} t^k.
\label{eqn:hilbshort}
\end{equation}
We recall that \(c_{k\, \br}\) is the number of independent operators which contain \(r_i\) powers of \(\ph_i\) fields and \(k\) derivatives, \emph{i.e.} operators schematically of the form \(\ph_1^{r_1}\dots\ph_N^{r_N}\pd^k\) where it is to be understood that the \(k\) derivatives act in some unspecified way on the \(\ph_1^{r_1}\dots\ph_N^{r_N}\).

The approach we take to computing \(H_N\) is as follows. We fix the number of $\phi$ fields (\emph{i.e.}, fix $\br$) and then count the number of derivatives we can add to form independent operators, that is, we consider the sum over $k$ in eq.~\eqref{eqn:hilbshort} first. For fixed $\br$ this is also a Hilbert series, which we denote by $H_\br(t)=\sum_{k=0}^{\infty} c_{k\,\br}t^k $ (where $N$ is left implicit). The full Hilbert series is regained by the sum over all $H_\br(t)$ weighted by $\bu^\br$:
\begin{equation}
H_N(t,\ui)=\sum_{\br=0}^{\infty} \bu^\br H_\br(t).
\end{equation}
A key advantage in fixing $\br$ is that IBP only relates operators of the same $\br$. Moreover, for fixed \(\br\) it makes sense to pass to Fourier space where implementing the equivalence under IBP and EOM becomes transparent.

We start by considering a single flavor and will then generalize the result to \(N\) flavors.
We perform a Fourier decomposition for the field \(\ph\), writing
\begin{equation}
\ph(\th) = \sum_{q=-\infty}^{\infty}a_qe^{iq\th},
\end{equation}
where the \(a_q\) are the Fourier coefficients. An operator composed of \(r\) powers of \(\ph\) fields and \(k\) derivatives
decomposes as
\begin{equation}
\ph^r\pd^k = \sum_{q_1,\dots,q_r=-\infty}^{\infty} w(q_1,\dots,q_r)a_{q_1}\dots a_{q_r} e^{i(q_1+\dots+q_r)\th},
\label{eqn:op_fourier}
\end{equation}
where \(w\) is a degree \(k\) polynomial in the momenta \(q_1,\dots,q_r\). One can always symmetrize $w$ due to the permutation symmetry among the dummy indices $q_i$---this is just the fact that the $\phi$'s are indistinguishable. Because a degree $k$ symmetric polynomial uniquely determines an operator with $k$ derivatives, and vice-versa,
we can translate the operator counting to that of counting polynomials. This leads us to consider {\it polynomial rings} in the  momenta; we shall see that
the IBPs and EOM imply polynomial constraints, the consequences of which are embodied in {\it ideals} of the rings.\footnote{Appendix~\ref{app:a} provides
a brief introduction to the concepts from commutative algebra which we use here.}

Let \(R_r=\mathbb{R}[q_1,\dots,q_r]\) be the polynomial ring in the \(r\) momenta with real coefficients. The symmetric polynomials form a subring \(R_r^{S_r}= \mathbb{R}[q_1,\dots,q_r]^{S_r}\subset R_r\). It is a well known result, \emph{e.g.}~\cite{Sturmfels:inv}, that the ring \(R_r^{S_r}\) is freely generated by the power sum symmetric polynomials \(P_1, \dots, P_r\) defined as
\begin{equation}
P_n = \sum_{s=1}^{r}q_s^n,
\label{eqn:power_sum}
\end{equation}
and so we have \(R_r^{S_r} = \mathbb{R}[P_1,\dots,P_r]\).

What happens when we consider IBP and EOM? In the above exposition, integration by parts manifests as the statement of momentum conservation. When an operator in the action, \(\int_{s^1} d\th \ph^r\pd^k\), is Fourier decomposed, integrating over \(\th\) forces \(P_1 = q_1 + \dots +q_r\) to vanish,
\begin{equation}
\int_{S^1} d\th\, e^{iP_1\th} = 2\pi \d_{P_1,0}.
\label{eqn:delta}
\end{equation}
From eq.~\eqref{eqn:op_fourier}, it is clear that a total derivative brings down a factor of \(P_1 = q_1 + \dots + q_r\). An operator is therefore a total derivative if and only if the polynomial \(w(q_1,\dots,q_r)\) is proportional to \(P_1\); hence, the equivalence relation in eq.~\eqref{eqn:equiv_IBP} is translated to \(w_a \sim w_b\) if \(w_a = w_b + w_cP_1\) for \(w_{a,b,c}\in R_r^{S_r}\). Algebraically, the statement of momentum conservation, \(P_1 = 0\), defines an ideal \(\avg{P_1}\) of \(R_r^{S_r}\). The set of operators containing  \(r\) powers of \(\ph\) fields modulo IBP lie in the quotient ring \(R_r^{S_r}/\avg{P_1}\).

In momentum space the EOM translates to \(q^2 = 0\), which implies that \(P_n = 0\) for \(n\ge 2\), as is obvious from eq.~\eqref{eqn:power_sum}. Additionally, the EOM imply \(P_1^{r+1}= (q_1 + \dots + q_r)^{r+1} = 0\), since every term in the expansion necessarily involves a $q_i^2$.   This embodies the fact that EOM only allow $r$ derivatives to be distributed onto $\phi^r$; application of a  further derivative necessarily requires a $\dd^2 \phi$ in the operator. To study the equivalence under the EOM we therefore examine the ideal of \(R_r^{S_r}\) generated by \(\avg{P_1^{r+1},P_2,\dots,P_r}\).

Taken together, IBP and the EOM define the ideal \(I_r = \avg{P_1,P_1^{r+1},P_2,\dots,P_r}\) and equivalence classes of operators lie in the quotient ring
\begin{equation}
\bigslant{R_r^{S_r}}{I_r} = \bigslant{\mathbb{R}[P_1,\dots,P_r]}{\avg{P_1,P_1^{r+1},P_2,\dots,P_r}} =\bigslant{\mathbb{R}[P_1]}{\avg{P_1,P_1^{r+1}}}.
\end{equation}
In the quotient ring, the EOM simply remove the generators \(P_2, \dots, P_r\). For the one flavor case, momentum conservation also removes the generator \(P_1\) and the quotient ring is trivial; it consists only of the identity element. The reason we do not indicate this in the above equation is that the above form is well suited for generalization to the \(N\) flavor case. For the single flavor case, \(R_r^{S_r}/I_r\) being trivial reflects the fact that we can use integration by parts and equations of motion to remove all operators with derivatives acting on \(\ph\)---any term of the form $\phi^{r-k} (\dd \phi)^k$ with $k>0$ can be written as a total derivative: $\dd( \phi^{r-k+1} (\dd \phi)^{k-1})$.

The generalization to \(N\) flavors is straightforward and follows the exact steps as the one flavor case. Each field is Fourier decomposed with the \(i\)th flavor having Fourier coefficients \(a_q^{(i)}\), \(i = 1, \dots, N\). An operator composed of \(\mathbf{r}\) powers of \(\mathbf{\ph}\) fields and \(k\) derivatives is decomposed in the action as,
\begin{align}
\int_{S_1} d\th \, \ph_1^{r_1}\dots \ph_{N}^{r_N} \pd^k = \sum_{\{q^{(1)}\}\dots \{q^{(N)}\} = -\infty}^{\infty} & w\big(\{q^{(1)}\}\dots \{q^{(N)}\}\big) \times \big(a_1^{(1)}\dots a_{r_1}^{(1)}\big) \dots \big(a_1^{(N)}\dots a_{r_N}^{(N)}\big) \nonumber \\
&\times \int_{S_1} d\th \exp\left( i \th \sum_{i=1}^N\sum_{j=1}^{r_i}q_j^{(i)}\right),
\end{align}
where \(w\) is a degree \(k\) polynomial of the momenta, invariant under the symmetric group \(S_{r_i}\) for each set of momenta \(\{q^{(i)}\}\). We denote this ring by \(R_{\mathbf{r}}^{G_{\br}} = \mathbb{R}[\{q^{(1)}\},\dots,\{q^{(N)}\}]^{G_{\br}}\) where \(G_{\br} = S_{r_1} \times \dots \times S_{r_N}\).  \(R_{\br}^{G_{\br}}\) is freely generated by the power sum symmetric polynomials \(P_1^{(i)},\dots,P_{r_i}^{(i)}\): \(R_{\br}^{G_{\br}} = \mathbb{R}[\{P^{(1)}\},\dots,\{P^{(N)}\}]\).

As in eq.~\eqref{eqn:delta}, IBP is handled by momentum conservation, \(P_1^{(1)} + \dots + P_1^{(N)} = 0\), while the EOM imply \(P_n^{(i)} = 0\) for \(n \ge 2\) as well as \(\big(P_1^{(i)}\big)^{r_i+1} = 0\). Together, these equations form an ideal of \(R_{\br}^{G_{\br}}\). The quotient ring containing the equivalence classes of operators is
\begin{equation}
\bigslant{R_{\br}^{G_{\br}}}{I_{\br}} = \bigslant{\mathbb{R}[P_1^{(1)},\dots,P_1^{(N)}]}{\big\langle P_1^{(1)}+\dots + P_1^{(N)},\big(P_1^{(1)}\big)^{r_1+1},\dots,\big(P_1^{(N)}\big)^{r_N+1}}\big\rangle.
\label{eqn:module}
\end{equation}

Questions about the EFT basis can now be studied by examining the structure of the modules in eq.~\eqref{eqn:module}. One of our main interests is determining the number of operators in the EFT basis, and this information is encoded in the Hilbert series \(H_{\br}(R_{\br}^{G_{\br}}/I_{\br},t)\).
Computing \(H_{\br}(R_{\br}^{G_{\br}}/I_{\br},t)\) for arbitrary \(\br\) and \(N\) is fairly involved. However, \(H_{\br}(R_{\br}^{G_{\br}}/I_{\br},t)\) can be easily obtained for any specific value of \(\br\) and \(N\) through the use of algebro-geometric computer packages such as {\tt Macaulay2}~\cite{m2} (see appendix~\ref{app:b} for an example). 

Another important aspect of this framework that we wish to emphasize is the ability to obtain an explicit basis of independent operators. This is simply the set of operators corresponding to the basis elements of the module, eq.~\eqref{eqn:module}. These are also easily output in a package such as {\tt Macaulay2}---see Appendix~\ref{app:b} for an example.

In the next two sections we will proceed to address the specifics of computing \(H_{\br}(R_{\br}^{G_{\br}}/I_{\br},t)\) in general, and obtain summed formulas
for the generalized Hilbert series, eq.~\eqref{eqn:Hdef}, for our one-dimensional EFT. We end this section with some observations that will be useful
to keep in mind during the following. First, due to the equations of motion, the quotient ring in eq.~\eqref{eqn:module} contains only a finite number of elements, and so
the full generalized Hilbert series will be finitely generated.
This is because the EOM bound the number of derivatives we can add to operator of the
form \(\ph_1^{r_1}\dots\ph_N^{r_N}\)---one obvious consequence being \(k_i\le r_i\), where \(k_i\) denotes the number of derivatives acting on \(\ph_i\). Second,
it is useful to think of  \(H_{\br}(R_{\br}^{G_{\br}}/I_{\br},t)\) heuristically having the form
\begin{equation}
H_{\br}(R_{\br}^{G_{\br}}/I_{\br},t) = \sum_{k=0}^{\infty}t^k \sum_{k_1+\dots+k_N = k}\big\{\text{conditions}\big\},
\label{eqn:heuristic_H_r}
\end{equation}
where \{conditions\} abstractly denotes the conditions for counting independent degree \(k\) polynomials in the module. Such conditions
 are encoded in the specific form of the ideal; for example, the consequence  \(k_i\le r_i\) noted above is reflected by the equations \(\big(P_1^{(i)}\big)^{r_i+1}\) in the ideal of eq.~\eqref{eqn:module}. Third,  because
 we sum over \(\br\) to get the full Hilbert series, \(H_N = \sum_{\br} \bu^{\br}H_{\br}\), and because,  in general, \(H_{\br}\) is a piecewise function of \(\br\),
  it may prove most prudent to leave \(H_{\br}\) as a sum formula of the form in eq.~\eqref{eqn:heuristic_H_r}. We will encounter such a situation
  when faced with the general formula for $H_N$, eq.~\eqref{eqn:HFinal_Sum}, the derivation of which we now turn to.


\section{Computing the Hilbert series $\hnui$} \label{sec:intuitive_calc}

The aim of this section is to obtain a formula for the generalized Hilbert series, eq.~\eqref{eqn:Hdef}, for the one-dimensional EFT of $N$ real scalars fields we have been considering. The final result for general $N$, presented as a sum formula, is given in eq.~\eqref{eqn:HFinal_Sum}. Rather than jumping straight from eq.~\eqref{eqn:module} to this result, we begin by considering three simpler cases---no relations, relations only from IBP, and relations only from use of the EOM---our aim being to show results which are intuitively easy to understand, as well as simple to obtain from the framework of the previous section. In this section we also emphasize the combinatorial interpretations of our results, which have a natural and well-studied connection with Hilbert series.

\subsection*{No relations}
In counting operators, the easiest place to begin is to not impose any relations. How many operators can be formed from derivatives acting on the \(\ph_i\), \(i=1,\dots,N\)?
In this case, the independent operators are monomials in \(\ph_i, \pd \ph_i, \pd^2 \ph_i, \pd^3 \ph_i, \dots\), \emph{i.e.} every operator is obtained in the expansion of
\begin{align*}
\prod_{i=1}^{N}\big(1+\ph_i + \ph_i^2 + \dots\big)\big(1+\pd\ph_i &+ (\pd\ph_i)^2 + \dots\big)\big(1+\pd^2\ph_i + (\pd^2\ph_i)^2 +\dots \big)\dots \\
&=\frac{1}{\prod_{i=1}^N(1-\ph_i)(1-\pd\ph_i)(1-\pd^2\ph_i)\dots} .
\end{align*}
We say that the operator basis is freely generated by the set of operators \(\{\pd^k \ph_i\}\), \(k = 0,\dots, \infty\).

The generating set of operators \(\{\pd^k\ph_i\}\) have corresponding weights \(\{t^ku_i\}\). Since there are no non-trivial relations among these generators, the Hilbert series is 
\begin{equation}
H_{N,\text{free}}(t,\ui) = \frac{1}{\prod_{i=1}^N (1-u_i)(1-tu_i)(1-t^2u_i)(1-t^3u_i)\dots} = \frac{1}{\prod_{i=1}^{N}(u_i;t)_{\infty}},
\label{eqn:H_free}
\end{equation}
where \((u;t)_{\infty} = \prod_{k=0}^{\infty}(1-t^ku)\) is the $q$-Pochhammer symbol and the subscript ``free'' denotes that we are not imposing any IBP or EOM relations in this counting. The $q$-Pochhammer symbol gives us a clear interpretation of the series coefficients in terms of partitions, which we return to at the end of this subsection.

Let us show how the above Hilbert series is obtained using the framework of section~\ref{sec:framework}. For clarity, we consider the \(N = 1\) case; generalization to arbitrary \(N\) is straightforward. For fixed \(r\), the module is \(R^{S_r} = \mathbb{R}[P_1,\dots,P_r]\), where the \(P_n\) carry weight \(t^n\). The number of degree \(k\) polynomials in \(R^{S_r}\) is number of points in the set \(\{(k_1,\dots,k_r)~|~k_1 + 2k_2 + \dots + rk_r = k\}\) so that the Hilbert series is 
\begin{equation}
H_{r,\text{free}}(R^{S_r},t) = \sum_{k=0}^{\infty}t^k\sum_{k_1+2k_2+\dots+rk_r = k} = \sum_{k_1=0}^{\infty}\dots \sum_{k_r=0}^{\infty} t^{k_1 + 2k_2 + \dots +rk_r} = \frac{1}{\prod_{n=1}^{r}(1-t^n)}.
\label{eqn:Hil_symm_poly}
\end{equation}
The above result is made transparent by recognizing that all independent monomials in \(R^{S_r}\) are obtained in the expansion of \((1+P_1 +P_1^2+\dots)(1+P_2+P_2^2+\dots)\dots(1+P_r+P_r^2+\dots)\); substituting \(P_n\) with its weight \(t^n\) and geometrically summing produces the above result. The full Hilbert series is
\begin{equation*}
H_{1,\text{free}}(t,u) = \sum_{r=0}^{\infty} \frac{u^r}{\prod_{n=1}^{r}(1-t^n)} = \frac{1}{(u;t)_{\infty}},
\end{equation*}
reproducing eq.~\eqref{eqn:H_free} for \(N=1\).

Let us now give an interpretation of the coefficients in the expansion of the Hilbert series, \(H_{N,\text{free}} = \sum_k \sum_{\br} c_{k\, \br, \text{free}} \bu^{\br}t^k\).
The number of operators composed of \(k\) derivatives acting on \(\br\) powers of \(\ph\) fields is the number of ways of partitioning the \(k\) derivatives onto \(\ph_1^{r_1}\dots \ph_N^{r_N}\). Specifically, for \(N=1\), \(c_{k\, r, \text{free}} = p(k;r)\) is the number of (indistinct) partitions of \(k\) into at most \(r\) parts.\footnote{A partition of \(n\) is a sequence of integers \(\l(n) = (\l_1,\l_2,\dots,\l_l)\) such that \(\l_1 + \dots + \l_l = n\) and the sequence weakly decreases, \(\l_i \ge \l_{i+1}\). The length of a partition is the number of non-zero \(\l_i\), \(\abs{\l(n)} = l\). The partitions of \(n\) into at most \(m\) parts is the set \(L(n;m) = \{ \l(n) | \sum_{i=1}^{\abs{\l(n)}}\l_i = k, \abs{\l(n)} \le m\}\); the cardinality of this set is \(\abs{L(n;m)}=p(n;m)\). For example, the partitions of 4 are \(\{(4),(3,1),(2,2),(2,1,1),(1,1,1,1)\}\) so that, \emph{e.g.}, \(p(4;2) = 3\). Obviously, if \(m\ge n\) then \(p(n;m) = p(n)\). By definition, \(p(0;0) = 1\) while \(p(n;0) =0\) for \(n\ge 1\).}
For general \(N\),
\[
c_{k r_1\dots r_N, \text{free}} = \sum_{k_1+\dots+k_N = k}p(k_1;r_1)\dots p(k_N;r_N),
\]
\emph{i.e.} \(c_{k\, \br,\text{free}}\) is the number of ways of distinctly partitioning \(k\) into \(N\) parts (\(k_1+ \dots + k_N = k\)) weighted by the number of indistinct partitions of the \(k_i\) into at most \(r_i\) parts.

\subsection*{Only relations from integration by parts}
Operators which are total derivatives vanish in the action, leading to relations from integration by parts. For simplicity, let us consider the one flavor case and count the number of operators composed of \(r\) powers of \(\ph\) fields and \(k\) derivatives, modulo integration by parts. Per the discussion above, the number of operators of the form \(\ph^r\pd^k\) is the number of partitions of \(k\) into at most \(r\) parts, \(p(k;r)\). Relations amongst the \(p(k;r)\) operators from IBP are formed from taking the \(p(k-1;r)\) operators with one less derivative and applying a total derivative. For example, \((\pd^4\ph)\ph^{r-1}\) and \((\pd^3\ph)(\pd\ph)\ph^{r-2}\) are related via \(\pd((\pd^3\ph)\ph^{r-1}) =0\).

The \(p(k-1,r)\) relations obtained in the above manner are all independent\footnote{One can see this by standing ordering scheme arguments, making using of the natural scheme induced by the weakly decreasing condition on partitions.}
and therefore the number of independent of operators modulo integration by parts is\footnote{Except in the case for \(k=1\), \(r=0\), where we have \(c_{1\, 0} = 0\).}
\begin{equation}
c_{k r, \text{IBP}} = p(k;r) - p(k-1;r),
\end{equation}
 and hence the Hilbert series is given by
\begin{equation}
H_{1,\text{IBP}}(t,u) = \frac{1}{(u;t)_{\infty}} - \frac{t}{(u;t)_{\infty}} + t.
\end{equation}
In the expansion of the second term above, the \(t\) in the numerator ensures that the coefficient of \(u^rt^k\) is \(p(k-1;r)\). The sole \(t\) in the above cancels the linear \(u^0t^1\) piece in the expansion of the second term. The straightforward generalization to \(N\) flavors is,
\begin{equation}
H_{N,\text{IBP}}(t,\ui) = \frac{1-t}{\prod_{i=1}^N(u_i;t)_{\infty}} + t.
\label{eqn:H_IBP}
\end{equation}

This result is also readily obtained using the formalism from section~\ref{sec:framework}. Let us consider the general \(N\) flavor case. For fixed \(r_1,\dots,r_N\), equivalence classes of operators related by integration by parts lie in the quotient ring
\begin{equation*}
\bigslant{R_{\br}^{G_{\br}}}{I_{\br}} = \bigslant{\mathbb{R}[\{P^{(1)}\},\dots,\{P^{(N)}\}]}{\avg{P_1^{(1)} + \dots + P_1^{(N)}}}.
\end{equation*}
The ideal reflects the statement of momentum conservation and accounts for integration by parts. In essence, this ideal removes one of the \(P_1^{(i)}\) generators from \(R_{\br}^{G_{\br}}\) when we construct the quotient ring. More precisely, when the ideal is defined by a single, homogeneous polynomial of degree \(j\), \(I = \avg{f}\) with \(\text{deg}(f) = j\), an elementary calculation tells us that the number of independent degree \(k\) polynomials in \(R/I\) is equal to the number at degree \(k\) in \(R\) minus the number at degree \(k-j\) in \(R\) (see appendix~\ref{app:a}). Therefore, since \(P_{1}^{(1)} + \dots + P_{1}^{(N)}\) is homogeneous and of degree one, the Hilbert series of the quotient ring is given by
\begin{align*}
H_{\br,\text{IBP}}(R_{\br}^{S_{\br}}/I_{\br};t) &= \sum_{k=0}^{\infty}t^k \left[\sum_{k_1+\dots+k_N = k} - \sum_{k_1+\dots+k_N = k-1}\right] \\
&= \frac{1-t}{\prod_{i=1}^N\prod_{n_i=1}^{r_i}(1-t^{n_i})},
\end{align*}
where the second equality holds for \(\br \ne 0\) and is unity for \(\br = 0\). Each set \(\{P^{(i)}\}\) contributes to the above Hilbert series analogous to the single set in eq.~\eqref{eqn:Hil_symm_poly}; the \(1-t\) in the numerator reflects the fact that the ideal essentially removes one of the \(P_1^{(i)}\). Summing \(\sum_{\br} \mathbf{u}^{\br}H_{\br}(t)\) we obtain the full Hilbert series as in eq.~\eqref{eqn:H_IBP}.

\subsection*{Only relations from equations of motion}
The equations of motion are \(\pd^2\ph_i = 0\); consequently \(\pd^k\ph_i = 0\) for \(k\ge 2\). Thus, when including only the relations from equations of motion, all operators are generated by the set \(\{\ph_i,\pd\ph_i\}\). Note that the EFT basis is finitely generated. There are no non-trivial relations amongst the generators, and therefore the Hilbert series is
\begin{equation}
H_{N,\text{EOM}}(t,\ui) = \frac{1}{\prod_{i=1}^N(1-u_i)(1-tu_i)}.
\label{eqn:H_EOM}
\end{equation}

In the language of section~\ref{sec:framework}, at fixed \(r_1,\dots,r_N\) we study the module
\begin{equation*}
R_{\br}^{G_{\br}}/I_{\br} = \bigslant{\mathbb{R}[P_1^{(1)},\dots,P_1^{(N)}]}{\big\langle \big(P_1^{(1)}\big)^{r_1+1},\dots,\big(P_1^{(N)}\big)^{r_N+1}}\big\rangle,
\end{equation*}
whose Hilbert series is given by
\begin{equation}
H_{\br,\text{EOM}}(R_{\br}^{G_{\br}}/I_{\br},t) = \sum_{k=0}^{\infty}t^k \sum_{k_1+\dots+k_N = k} I_{\{k_i \le r_i\}}.
\label{eqn:H_r_EOM}
\end{equation}
Here we have adopted a notation $I_A$, whose value is $1$ if the condition $A$ is satisfied, and $0$ otherwise. It is also understood that $i$ runs through $1,\cdots,N$ in the set of conditions $\left\{k_i \le r_i\right\}$. The EOM imply the constraints \(k_i\le r_i\) and are directly seen from the \(\big(P_1^{(i)}\big)^{r_i+1}\) terms in the ideal. We can geometrically sum the above to obtain
\begin{equation}
H_{\br,\text{EOM}}(R_{\br}^{G_{\br}}/I_{\br},t) = \sum_{k_1=0}^{r_1}\dots \sum_{k_N=0}^{r_N} t^{k_1+\dots+k_N}= \frac{\prod_{i=1}^{N}(1-t^{r_i+1})}{(1-t)^N}.
\label{eqn:H_r_EOM_sum}
\end{equation}
We note that the reason why this sum was simple is algebraically rooted in the fact that each term in the ideal depends only on a single indeterminate \(P_1^{(i)}\).
Upon summing, we reproduce eq.~\eqref{eqn:H_EOM}:
\begin{align}
H_{N,\text{EOM}}(t,\ui) &= \sum_{r_i\ge0}\sum_{k\ge0}\sum_{k_1+ \dots + k_N = k} u_1^{r_1}\dots u_N^{r_N} t^k I_{\{k_i\le r_i\}} \nonumber \\
&= \sum_{r'_i\ge 0}u_1^{r'_1}\dots u_N^{r'_N}\sum_{k_i\ge 0} (tu_1^{})^{k_1}\dots (tu_N^{})^{k_N} \nonumber \\
&= \frac{1}{\prod_{i=1}^N(1-u_i)(1-tu_i)},
\end{align}
where in the first to second line we defined \(r_i' = r_i-k_i\). We have explicitly written out this step to highlight a point made at the end of section~\ref{sec:framework}: in order to make use of sum manipulations when computing the full Hilbert series, it is frequently simpler to leave \(H_{\br}\) as a sum formula---as in eq.~\eqref{eqn:H_r_EOM}---rather than first finding a closed form sum for \(H_{\br}\)---as in eq.~\eqref{eqn:H_r_EOM_sum}. While this distinction is mild for the present case, it is quite useful for the sums considered in the next subsection.

\subsection*{Relations from both integration by parts and equations of motion}
We now turn to the task of computing the Hilbert series when we account for relations from both integration by parts and the equations of motion. As we will see, the Hilbert series in this case is much more interesting than when these relations are considered independently. Unlike the three previous cases, the generators of the EFT basis are not so easy to guess and there are non-trivial relations among them, leading to a rich structure in the Hilbert series.

Our launching point is the quotient ring of eq.~\eqref{eqn:module}, reproduced here for convenience, which describes equivalence classes of operators at fixed \(\br\)
\begin{equation*}
\bigslant{R_{\br}^{G_{\br}}}{I_{\br}} = \bigslant{\mathbb{R}[P_1^{(1)},\dots,P_1^{(N)}]}{\big\langle P_1^{(1)}+\dots + P_1^{(N)},\big(P_1^{(1)}\big)^{r_1+1},\dots,\big(P_1^{(N)}\big)^{r_N+1}}\big\rangle.
\end{equation*}
We wish to find a sum formula for the Hilbert series of this module, schematically of the form
\begin{equation*}
H(R_{\br}^{G_{\br}}/I_{\br},t) = \sum_{k=0}^{\infty}c_{k\, \br} t^k \sim \sum_{k=0}^{\infty}t^k \sum_{k_1+\dots+k_N = k}\big\{\text{conditions}\big\}.
\end{equation*}
Instead of resorting to involved mathematics, to obtain the coefficients \(c_{k\, \br}\) we build on the experience gained from studying the previous simpler systems. Due to the EOM, the \(\big(P_1^{(i)}\big)^{r_i+1}\) terms in the ideal require \(k_i \le r_i\), as in eq.~\eqref{eqn:H_r_EOM}. As when we handled IBP alone, an independent IBP relation among the $\phi_1^{r_1} \cdots \phi_N^{r_N} \partial^k$ operators is generated from a total derivative acting on each of the operators with one less derivative.
In this spirit, it is very tempting to write
\begin{equation*}
c_{k\, \textbf{r}}^{} \sim c_{k\, \textbf{r},\text{EOM}}^{} - c_{k-1\, \textbf{r},\text{EOM}}^{} =\sum_{k_1+\dots+k_N = k}I_{\{k_i\le r_i\}} - \sum_{k_1+\dots+k_N = k-1}I_{\{k_i\le r_i\}}.
\end{equation*}
However, it is almost immediately obvious that this expression cannot be correct, because it will go negative when $k$ is too large. When this happens, we should take $c_{k\, \textbf{r}}=0$. It turns out that the correction condition to guarantee \(c_{k\, \br}\) be non-negative is $2k\le r\equiv r_1+\cdots +r_N$. Therefore, we have
\begin{equation*}
c_{k\, \textbf{r}}^{} \propto I_{2k \le r} \left( c_{k\, \textbf{r},\text{EOM}}^{} - c_{k-1\, \textbf{r},\text{EOM}}^{} \right).
\end{equation*}

Additionally, there is another constraint coming from the interplay of IBP and EOM. Using IBP, one can always get rid of all derivatives acting on a certain field \(\ph_i\), reallocating all of the \(k\) derivatives onto the other fields. From the ideal, this is seen by using the momentum conservation equation \(\sum_{i=1}^NP_1^{(i)} = 0\) to eliminate a chosen \(P_1^{(i)}\).\footnote{Another way to see this is that \(\sum_{i=1}^NP_1^{(i)}\) remains in the Gr\"obner basis of the ideal \(I_{\br}\). The Hilbert series can be computed from the ideal generated by the initial monomials of the Gr\"obner basis. In a monomial order where \(P_1^{(i)} > P_1^{(j)}\) for all \(j \ne i\), we have \(\text{in}(\sum_{i=1}^NP_1^{(i)}) = P_1^{(i)}\) and hence all monomials involving \(P_1^{(i)}\) are eliminated from \(R_{\br}^{G_{\br}}/\text{in}(I_{\br})\). See appendix~\ref{app:a}. \label{foot:elim_deriv}} It is then clear that if $k$ is greater than $r-r_i$ for any $i \in \{1,\cdots,N\}$, the operator would be zero. Therefore, we also have
\begin{equation*}
c_{k\,\textbf{r}}^{} \propto I_{\left\{ {k \le r - {r_i}} \right\}} .
\end{equation*}

Combining these constraints, the Hilbert series is given by
\begin{equation}
H = \sum\limits_{\br=0}^\infty  \bu^\br \sum\limits_{k = 0}^\infty  t^k I_{\left\{ {k \le r - {r_i}} \right\}}I_{2k \le r}\left[ \sum\limits_{{k_1} +  \cdots  + {k_N} = k} I_{\left\{ {{k_i} \le {r_i}} \right\}} - \sum\limits_{{k_1} +  \cdots  + {k_N} = k - 1} I_{\left\{ {{k_i} \le {r_i}} \right\}} \right] . \label{eqn:HFinal_Sum}
\end{equation}
Note that the conditions \(I_{\{k \le r - r_i\}}\) and \(I_{2k \le r}\) have overlap---which one dominates depends on \(r_1,\dots,r_N\).
Let \(r_j\) be the maximum number in a given \(r_1,\dots,r_N\), \(r_j \ge r_i\) for all \(i\). The set of conditions \(\{k \le r -r_i\}\) is then equivalent to the single condition \(k \le r - r_j\). If \(r_j \ge \sum_{i\ne j}r_i\), then the condition \(2k \le r\) is unnecessary---in this situation \(H_{\br}(R_{\br}^{G_{\br}}/I_\br,t)\) is relatively simple to compute.  If \(r_j < \sum_{i\ne j}r_i\), then it is necessary to include \(I_{2k \le r}\) while the conditions \(\{k\le r - r_i\}\) are automatically satisfied.

With eq.~\eqref{eqn:HFinal_Sum}, we can obtain the closed form of the Hilbert series for relatively low numbers of flavors, $N \le 3$
\begin{eqnarray}
H_1 &=& \frac{1}{{1 - {u_1}}} , \nonumber\\
H_2 &=& \frac{1}{{\left( {1 - {u_1}} \right)\left( {1 - {u_2}} \right)\left( {1 - t{u_1}{u_2}} \right)}} , \label{eq:someres} \\
H_3 &=& \frac{{1 - t{u_1}{u_2}{u_3}}}{{\left( {1 - {u_1}} \right)\left( {1 - {u_2}} \right)\left( {1 - {u_3}} \right)\left( {1 - t{u_1}{u_2}} \right)\left( {1 - t{u_1}{u_3}} \right)\left( {1 - t{u_2}{u_3}} \right)}} \nonumber.
\end{eqnarray}
For larger $N$, it becomes very laborious to directly sum up the expression eq.~\eqref{eqn:HFinal_Sum}. 
Moreover, the form of the Hilbert series becomes increasingly complicated. For example, for $N=4,5$ we have
\be
H_4 &=& \frac{1-t (s_3-s_4) - t^2(s_4 -s_1s_4)-t^3 s_4^2}{\prod_i (1-u_i) \,\prod_{i<j} (1-tu_iu_j)}\,,\nonumber   \\
H_5 &=& \frac{1}{\prod_i (1-u_i) \,\prod_{i<j} (1-tu_iu_j)}\bigg[ 1-t (s_3-s_4+s_5) - t^2(s_4 -s_1s_4+s_1s_5)  \label{eq:someres2}\\
&~&\!\!\!\!\!\!\!\!\!\!\!\!\!\!\!\!\!\!\!\!\!\!\!\!\!-t^3(s_4^2-s_1s_5+s_1^2s_5-s_2s_5-s_3s_5-s_4s_5) - t^4(s_4s_5 -s_1s_4s_5+s_1s_5^2) - t^5(s_5^2 -s_1s_5^2+s_2s_5^2) +t^6 s_5^3 \bigg] \,, \nonumber
\ee 
where $s_m$ are the elementary symmetric polynomials in \(N\) variables \(u_1,\dots,u_N\), 
\[s_m = \sum_{1\le i_1<\dots<i_m \le N}u_{i_1}\dots u_{i_m},\] (and where the value of $N$ for $s_m$ is left implicit in eqs.~\eqref{eq:someres2}). These can 
be more readily obtained via residues that the sum formula eq.~\eqref{eqn:HFinal_Sum} can be shown to have (see section~\ref{sec:poles} for further discussion on this point), or through the connection with $SL(2,{\mathbb C})$ representation theory that this EFT enjoys, which is the subject to which we now turn.


\section{The result for general $N$ from $SL(2,{\mathbb C})$}
\label{sec:su2}
It turns out that our one-dimensional theory with scalars has an interesting connection with \(SL(2,{\mathbb C})\) representation theory\footnote{We are grateful to Yuji Tachikawa for pointing out this connection to us.} that allows us to obtain a closed form of the Hilbert series 
relatively simply for an arbitrary number of flavors. As discussed in the last section, when we only include relations from equations of motion, all operators are generated by the set \(\{\ph_i, \pd \ph_i\}\). Pairing \(\ph_i\) and \(\pd \ph_i\) together into a doublet, the derivative acts as a lowering operator since \(\pd^2\ph_i = 0\). This is the origin of an underlying \(SL(2,{\mathbb C})\) structure which can be seen acting on the complex weights we use to construct the Hilbert series.

The \(SL(2,{\mathbb C})\) structure can be made manifest in the Hilbert series by a simple change of variables on the weights: rescale \(u_i \to \a \bau_i\) and \(t \to 1/\a^2\). For example, the single flavor Hilbert series with just equations of motion, eq.~\eqref{eqn:H_EOM}, becomes
\begin{subequations}
\begin{align}
\frac{1}{(1-u)(1-tu)} &\to \frac{1}{(1-\bau\a)(1-\bau\a^{-1})} \\
&= 1 + \bau(\a + \a^{-1}) + \bau^2(\a^2 + 1 + \a^{-2}) + \bau^3(\a^3 + \a + \a^{-1} + \a^{-3}) + \dots.
\end{align}
\end{subequations}
We immediately recognize the coefficient of \(\bau^r\) in the above expansion to be the character \(\ch_{r+1}(\a)\) of the \(r+1\) dimensional representation of \(SL(2,{\mathbb C})\): this Hilbert series is the character generating function for \(SL(2,{\mathbb C})\), \(H_{1,\text{EOM}}(\a;\bau) = \sum_{r=0}^{\infty}\bau^r\ch_{r+1}(\a)\).
The important observation is that for each irreducible representation of $SL(2,{\mathbb C})$, the terms not of highest weight ({\it i.e.} the coefficients of $\bau^r$ other than
 $\alpha^r$) can be obtained via application of the lowering operator---the derivative. Thus, if we want to account for IBPs, we simply disregard these
 terms in the series, since they correspond to total derivatives. By inspection, we find \(H_{1}(\a,\bau) = \sum_{r=0}^{\infty}\bau^r \a^r =1/(1-\bar{u}\a) \), reproducing our previous result for the $N=1$ Hilbert series (see eq.~\eqref{eq:someres}).

How does this picture generalize to arbitrary $N$? As is clear from eq.~\eqref{eqn:H_EOM}, the Hilbert series from just EOM for \(N\) flavors is simply \(N\) copies of the character generating function,
\begin{subequations}
\begin{align}
H_{N,\text{EOM}}(\a;\{\bau_i\}) &= H_{1,\text{EOM}}(\a;\bau_1)\dots H_{1,\text{EOM}}(\a;\bau_N) \\
&= \sum_{r_1,\dots,r_N}\bau_1^{r_1}\dots \bau_N^{r_N} \ch_{r_1+1}(\a)\dots\ch_{r_N+1}(\a) .
\end{align}
\label{eq:31}
\end{subequations}
We next perform a tensor decomposition, which expressed through the characters is written as
\be
 \ch_{r_1+1}(\a)\dots\ch_{r_N+1}(\a)= \sum_{k=0}^{\infty} \mathcal{C}^k_{\br}\ch_{k+1}(\a)  \,,
 \label{eq:tendec}
\ee
(for $N=2$ this is simply a Clebsch-Gordan decomposition), and we again make the observation that to convert the Hilbert series of eq.~\eqref{eq:31} into one taking IBPs into account, we  discard all but the highest weight of each irreducible representation on the rhs of eq.~\eqref{eq:tendec} (\textit{i.e.} put $\ch_{k+1}(\a)\to\a^k$). Thus we find
\be
H_{N}(\a,\{\bau_i\}) =  \sum_{r_1,\dots,r_N,k}\bau_1^{r_1}\dots \bau_N^{r_N} \alpha^k \,\mathcal{C}^k_{\br} \,,
\ee
which can be compared to the formula eq.~\eqref{eqn:hilbshort} in the original variables $(u,t)$, giving an alternative interpretation of the coefficient $c_{k\,\br}$  in terms of the tensor decomposition of eq.~\eqref{eq:tendec}: $c_{k\,\br}=\mathcal{C}^{|\br|-2k}_{\br}$.

\begin{figure}[t]
\centering
\begin{tikzpicture}[>=latex]
\node at (0,0) {\def\arraystretch{1.3}\(\displaystyle \left(\begin{array}{c} \ph_1 \\ \pd \ph_1 \end{array}\right) \otimes \left(\begin{array}{c} \ph_2 \\ \pd \ph_2 \end{array}\right) = \left( \begin{array}{c} \ph_1\ph_2 \\ \ph_1 \pd\ph_2 + \ph_2\pd\ph_1 \\ \pd\ph_1\pd\ph_2 \\  \ph_1 \pd\ph_2 - \ph_2\pd\ph_1 \end{array}\right)\)};

\draw[-,thick,dashed] (0.7,-.67)--(3.4,-.67);

\draw[-,thick,<-,blue] (3.3,0.9)--(5,0.9);
\draw[-,thick,<-,blue] (3.5,-0.9)--(5,0.8);

\draw[-,thick,<-,red] (3.2,-0.4)--(5,-0.9);
\draw[-,thick,<-,red] (3.4,0.2)--(5,-0.8);

\node (A) at (6.35,0.85) {Highest weights};
\node (B) at (6.6,-0.85) {Vanish by IBP/EOM};

\node (C) at (-3,-1.8) {$\bf 2$};
\node (D) at (-2,-1.8) {$\otimes$};
\node (E) at (-1,-1.8) {$\bf 2$};
\node (F) at (0.05,-1.8) {$=$};
\node (G) at (2,-1.8) {${\bf 3} ~~\oplus~~ {\bf 1}$};

\end{tikzpicture}
\caption{Decomposition of operator space into irreducible \(SL(2,\mathbb{C})\) representations.}
\label{fig:tendec}
\end{figure}

Let us summarize more precisely the above mapping onto \(SL(2,{\mathbb C})\) representation theory. With just EOM, operators are formed from polynomials in the \(\ph_i\) and \(\pd \ph_i\), {\it i.e.} they lie in the polynomial ring \(F_{\text{EOM}} = \mathbb{R}[\{\ph_i\},\{\pd\ph_i\}]\). Including integration by parts, we wish to find operators that are equivalent up to a total derivative; such operators lie in the space \(F_{\text{EOM}}/\pd F_{\text{EOM}}\). By \(SL(2,{\mathbb C})\) representation theory, this is equivalent to enumerating the irreducible representations contained in \(N\)-fold tensor products \(V_{r_1+1} \otimes \dots \otimes V_{r_N+1}\), where \(V_{r_i+1}\) is the \(SL(2,{\mathbb C})\) irreducible representation of dimension \(r_i+1\).

We now turn to computing the closed form of the Hilbert series for general \(N\).
This treatment makes reference to Molien's formula and the Haar integration measure, a general discussion of which we refer the reader to, \textit{e.g.}, the physics papers \cite{Gray:2008yu,Lehman:2015via}. For the following analysis to apply, we restrict $\alpha$ to lie in the maximal compact subgroup of $SL(2,{\mathbb C})$, namely $SU(2)$.

When converting the EOM result to include IBP, we discarded all terms which did not correspond to the highest weights in the $SL(2,{\mathbb C})$ irreducible representations. We can achieve this result directly by making use of the orthogonality of characters.\footnote{For a Lie group \(G\),  \(\int d\m_G^{}\, \chi_{r'}^*(g)\chi_{r}(g) = \d_{r'r}\) where \(\chi_r\) is the character in the \(r\)th representation and the integration is over all elements \(g\in G\) with \(d\m_G^{}\) the Haar measure. Since \(\ch_r(g) = \text{Tr}_r(g)\) is a class function, \(\ch_r(g) = \ch_r(h^{-1}gh)\) for \(g,h\in G\), the integration can be restricted to the maximal torus of \(G\). For \(G = SU(2)\) the maximal abelian subgroup is \(U(1)\), so the integral is over a single parameter \(\a\). In a somewhat sloppy notation, we write the Haar measure as \(d\m_{SU(2)}^{}(\a)\). For further discussions see~\cite{Gray:2008yu,Lehman:2015via}. Finally, for \(SU(2)\) we note that \(\chi_r^* = \chi_r\) since \(SU(2)\) is pseudo-real.} Specifically, to determine the multiplicity \(\mathcal{C}_{\br}^{r_0}\) for which the \((r_0+1)\)-dimensional representation of \(SU(2)\) appears in the tensor product \(V_{r_1+1}\otimes \dots \otimes V_{r_N+1}\), we multiply eq.~\eqref{eq:tendec} by \(\chi_{r_0+1}(\a)\) and integrate over \(d\m_{SU(2)}^{}(\a)\). To determine all such multiplicities in a given tensor product, we sum over \(r_0\). Applying this to eq.~\eqref{eq:31} and weighting the multiplicities by $\bau_0^{r_0}$, which plays the role of the highest weight of \(V_{r_0+1}\), we obtain \(H_N\),
\be
H_{N}(\bau_0,\{\bau_i\}) = \int d\mu_{SU(2)}^{}(\a) \,\sum_{r_0=0}^{\infty} \bar{u}_0^{r_0} \,\ch_{r_0+1}(\a) \,\,H_{N,\text{EOM}}(\a;\{\bau_i\}) \,.
\ee
Summing $\sum_{r_0=0}^{\infty} \bar{u}_0^{r_0}\,\ch_{r_0+1}(\a)   = H_{1,\text{EOM}}(\a; \bau_0)$, we find
 \be
H_{N}(\bau_0,\{\bau_i\}) &=& \int d\mu_{SU(2)}(\a) \,\, H_{1,\text{EOM}}(\a;\bau_0) H_{N,\text{EOM}}(\a;\{\bau_i\})   \nonumber \\
 &=& \int d\mu_{SU(2)}(\a) \,\, \prod_{i=0}^{N} H_{1,\text{EOM}}(\a; \bau_i) \, .
\label{eq:molein}
\ee
The above is Molien's formula applied to \(N+1\) \(SU(2)\) doublets, \textit{i.e.} \(H_N(\bau_0,\dots,\bau_N)\) may also be interpreted as counting the number of independent \(SU(2)\) invariants formed from \(N+1\) doublets! To make contact with the common notation for Molien's formula, as used in~\cite{Hanany:2010vu,Benvenuti:2006qr,Feng:2007ur,Gray:2008yu,Lehman:2015via,Sturmfels:inv}, note that
\begin{equation}
H_{1,\text{EOM}}(\a;\bau_i) = \frac{1}{(1-\bau_i \a)(1- \bau_i \a^{-1})} = \frac{1}{\text{det}(1-\bau_i g)} \, \label{eq:molein_notation}
\end{equation}
where \(g \in G\) is in the doublet representation of \(SU(2)\) and \(\a = e^{i\th}\) parameterizes the \(U(1)\) subgroup whose two-dimensional representation is \(e^{i\th \s_3}\) with \(\s_3\) the Pauli matrix. We note that the application of Molien's formula to $SU(2)$ invariants has also been treated in \cite{Gray:2008yu}.

To obtain the closed form of $H_{N}(\bau_0,\{\bau_i\})$ one evaluates Molien's formula in eq.~\eqref{eq:molein} by performing the contour integral specified by the Haar measure $\int d\mu_{SU(2)}(\a) = \oint_{|\alpha|=1} \frac{1}{2\pi i} \frac{d\a}{\a} \frac{1}{2} (1-\alpha^2) (1-\alpha^{-2})$. The pole structure is very simple, and the residue theorem can be applied easily for general $N$, with the result for $N\ge2$
\be
&~&{H_N}\left( {{\bau_0}, {\bau_1}, \cdots ,{\bau_N}} \right)  =  - \frac{1}{2}\sum\limits_{k = 0}^N {\frac{{\left( {1 - \bau_k^2} \right)\bau_k^{N - 2}}}{{\prod\limits_{i \ne k} {\left[ {\left( {1 - {\bau_k}{\bau_i}} \right)\left( {{\bau_k} - {\bau_i}} \right)} \right]} }}}  \nonumber \\
&~& =  - \frac{1}{2}\frac{1}{{\prod\limits_{i < j} {\left( {1 - {\bau_i}{\bau_j}} \right)} }}\frac{1}{{\prod\limits_{i < j} {\left( {{\bau_j} - {\bau_i}} \right)} }}\sum\limits_{k = 0}^N {\left\{ {{{\left( { - 1} \right)}^{N - k}}\left( {1 - \bau_k^2} \right)\bau_k^{N - 2}\prod\limits_{i < j \ne k} {\left[ {\left( {1 - {\bau_i}{\bau_j}} \right)\left( {{\bau_j} - {\bau_i}} \right)} \right]} } \right\}}  \,.\nonumber \\
\label{eqn:HN1}
\ee
The term in curly brackets is an antisymmetric polynomial in the $\bau_i$ $(i=0,\dots,N)$ and therefore divisible by $\prod_{i<j}\left(\bau_j-\bau_i\right)$. Hence, the quotient is a fully symmetric polynomial, as required by the symmetry of eq.~\eqref{eq:molein}. A few explicit values for low $N$ are
\be
H_1(\bau_0,\bau_1) &=& \frac{1}{(1-\bau_0\bau_1)} \nonumber  \,,\\
H_2(\bau_0,\bau_1,\bau_2) &=& \frac{1}{(1-\bau_0\bau_1)(1-\bau_0\bau_2)(1-\bau_1\bau_2)}  \,, \label{eqn:HN_ubar}\\
H_3(\bau_0,\bau_1,\bau_2,\bau_3) &=& \frac{1-\bau_0\bau_1\bau_2\bau_3}{(1-\bau_0\bau_1)(1-\bau_0\bau_2)(1-\bau_0\bau_3)(1-\bau_1\bau_2)(1-\bau_1\bau_3)(1-\bau_2\bau_3)} \,.  \nonumber
\ee
One can check that eqs.~\eqref{eq:someres}-\eqref{eq:someres2} are reproduced when sending $\bau_0\to t^{-1/2}$, $\bau_i\to u_i t^{1/2}$. 

The terms occuring in the Hilbert series in eqs.~\eqref{eqn:HN1}-\eqref{eqn:HN_ubar} have a simple interpretation from our understanding that \(H_N\) computes \(SU(2)\) invariants formed from \(N+1\) doublets. Denote these doublets by \(Q_{i\a}\) with \(i=0,\dots,N\) a flavor index and \(\a = 1,2\) a \(SU(2)\) index. The fundamental invariants are constructed from pairs of the \(Q_{i\a}\), \(M_{ij} = \e^{\a\b}Q_{i\a}Q_{j\b}\). All other \(SU(2)\) invariants can be formed from products of the \(M_{ij}\). That the \(M_{ij}\) generate \(SU(2)\) invariants is reflected by \(N+1 \choose 2\) terms of the form \((1-\bau_i\bau_j)\) in the denominator of the Hilbert series. As the \((N+1) \times 2\) matrix \(Q_{i\a}\) is at most rank two, the anti-symmetric matrix \(M_{ij}\) is at most rank two and therefore subject to relations among its components. These constraints take the explicit form \(0=M \wedge M = \e^{i_0,\dots,i_N}M_{i_0i_1}M_{i_2i_3}\).\footnote{Some readers may find these constraints more familiarly recognized as the Schouten identities, \(0 = \avg{ij}\avg{kl} - \avg{ik}\avg{jl} + \avg{il}\avg{jk}\).} These constraints, as well as the non-trivial relations among them (syzygies), govern the numerators of the Hilbert series. The simplest example is \(H_3\), where the single constraint \(0 = \e^{ijkl}M_{ij}M_{kl}\) is reflected by \(1 - \bau_0\bau_1\bau_2\bau_3\) in the numerator.

The poles of eq.~\eqref{eqn:HN1} occur at $\bau_i\bau_j=1$. The residues are easy to compute using the first line of eq.~\eqref{eqn:HN1},
\be
{\left. {\left( {1 - {\bau_a}{\bau_b}} \right){H_N}} \right|_{{\bau_a}{\bau_b} \to 1}} =  \frac{1}{{\prod\limits_{i \ne a,b} {\left[ {\left( {1 - {\bau_a}{\bau_i}} \right)\left( {1 - {\bau_b}{\bau_i}} \right)} \right]} }} \,.
\label{eqn:HNres}
\ee
Upon relabelling $\bau_a=1/\bau_b=\alpha$, we see this residue is in fact $H_{N-1,\text{EOM}}$. We will return to discuss further the pole structure of the Hilbert series in the following section.

Finally we note that this picture also provides a constructive method for finding the operators in the operator basis---they are the highest weight states of each irreducible representation obtained from decomposing the tensor products. A simple example is shown in Fig.~\ref{fig:tendec}, where one finds that the operator bases containing one power of $\phi_1$ and one power of $\phi_2$ are $\{\phi_1 \phi_2, \phi_1\partial\phi_2\}$ (for the second operator we have used equivalence under IBP: $ \phi_1\partial\phi_2\sim  \phi_1\partial\phi_2- \phi_2\partial\phi_1$). Other operators in the basis are clearly obtainable following an iterative procedure.


\section{Consistency conditions and relations among Hilbert series}
\label{sec:poles}

\begin{figure}[t]
\centering
\begin{tikzpicture}[>=latex]
\begin{scope}[scale=1.]
\node (A) at (-5.,3) {\Large \(H_{N,\text{EOM}}\)};
\node (B) at (5.,3) {\Large \(H_{N+1,\text{EOM}}\)};
\node (C) at (-5.,-3) {\Large \(H_N\)};
\node (D) at (5.,-3) {\Large \(H_{N+1}\)};

\draw[-latexnew,arrowhead=3mm,line width = 1pt] (-3.6,2.9) to (3.6,2.9);
\draw[-latexnew,arrowhead=3mm,line width = 1pt] (3.6,3.2) to (-3.6,3.2);

\node at (0,3.45) {\small \(\bou_{N+1} = 0\)};
\node at (0,2.55) {\small \(\displaystyle H_{N,\text{EOM}}(\a;\bou_1,\dots,\bou_N) H_{1,\text{EOM}}(\a;\bou_{N+1})\)};

\draw[-latexnew,arrowhead=3.mm,line width = 1pt] (3.6,-2.9) to (-3.6,-2.9);
\draw[-latexnew,arrowhead=3.mm,line width = 1pt] (-3.6,-3.2) to (3.6,-3.2);

\node at (0,-2.65) {\small \(\bou_{N+1} = 0\)};
\node[] at (0,-3.8) {\small \(\displaystyle \oint \frac{dx}{2\pi i} \frac{1}{x} H_{N}(\bou_0,\dots,\bou_{N-1},x)H_2(x^{-1},\bou_N,\bou_{N+1})\)};

\definecolor{mag2}{RGB}{255,0,255};

\draw[-latexnew,arrowhead=3.5mm,line width = 1pt] (-5.2,2.) to (-5.2,-2);

\node[rotate=90] at (-6.4,0) {\small \(\displaystyle \int d\m_{\a}\, H_{N,\text{EOM}}(\a;\bou_1,\dots,\bou_N)H_{1,\text{EOM}}(\a;\bou_0)\)};

\definecolor{rescol}{RGB}{17,119,51};

\draw[-latexnew,arrowhead=3.mm,line width = 1pt] (3.6,-2.) to (-3.6,2.);

\node[rotate=-29] at (0.2,0.2) {\small residue: \(\bou_i\bou_j \to 1\)};

\end{scope}
\end{tikzpicture}
\caption{The analytic structure of the Hilbert series for the EFT of real scalar fields in $d=1$.}\label{fig:hilb_conn}
\end{figure}
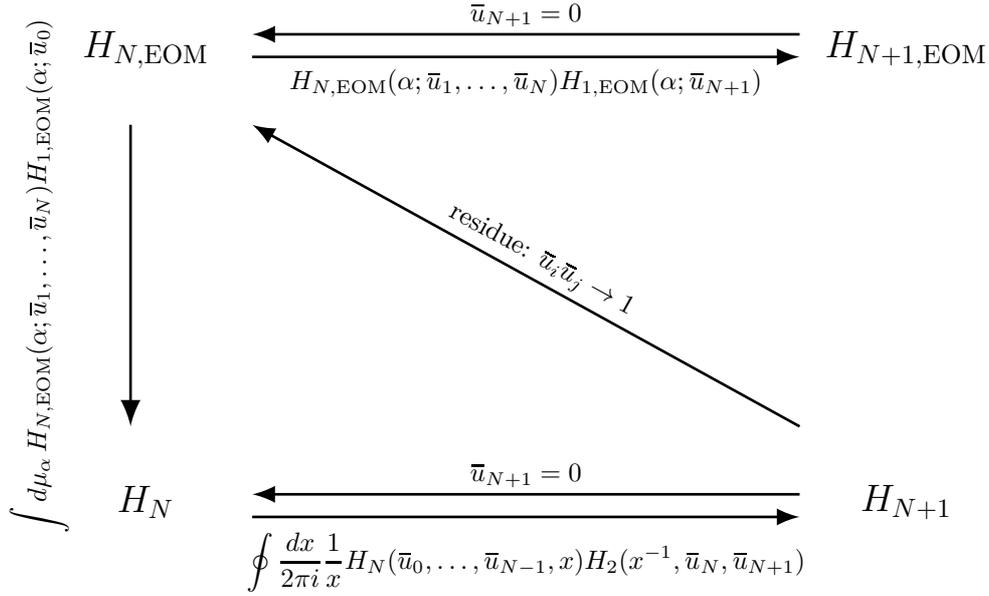

In the previous section we  presented a complementary method for studying operator bases in our one-dimensional  theory using $SL(2,{\mathbb C})$ representation theory that allowed us to obtain a closed form for the Hilbert series, with and without IBP, for general $N$. We now turn to analyzing these Hilbert series. Their various limits and analytic properties reveal  interesting connections between the different Hilbert series, summarized in Fig.~\ref{fig:hilb_conn}. We primarily work with the barred weights introduced in Sec.~\ref{sec:su2}, where these connections become more transparent. 
For convenience, we reproduce here the Hilbert series when only including relations from EOM (\(H_{N,\text{EOM}}\)) and including both EOM and IBP (\(H_N\)):
\begin{eqnarray}
H_{N,\text{EOM}}(\bou_0;\bou_1,\dots,\bou_N) &=& \frac{1}{\prod_{i= 1}^N(1- \bou_0\bou_i)(1-\bou_i/\bou_0)}, \label{eqn:HN_EOM_disc}\\
 H_N(\bou_0,\dots,\bou_N) &=& \frac{f(\bou_0,\dots,\bou_N)}{\prod\limits_{0\le i<j\le N}(1-\bou_i\bou_j)}, \label{eqn:HN_disc}
\end{eqnarray}
where \(f(\bou_0,\dots,\bou_N)\) is a symmetric polynomial in the \(\{\bou_i\}\) that can be explicitly determined from eq.~\eqref{eqn:HN1}. We note that \(H_N\) is symmetric in the \(\{\bou_i\}\), while \(\bou_0\) plays a separate role in \(H_{N,\text{EOM}}\); as in Sec.~\ref{sec:su2}, we occasionally use the notation \(\bou_0 = \a\) in \(H_{N,\text{EOM}}\) to distinguish this role. Lastly, we recall that the original \(t\) and \(\{u_i\}\) weights are re-expressed in terms of the \(\{\bou_i\}\) as \(t = 1/\bou_0^2\) and \(u_i = \bou_0\bou_i\).

We now examine various limits and residues of these Hilbert series. In the limit \(\bou_N \to 0\) of \(H_N\), we obtain the Hilbert series with one less flavor, \(H_{N-1}\). This clearly applies to  \(H_{N,\text{EOM}}\) as well. The poles of \(H_N\) occur at \(\bou_a\bou_b \to 1\) and their residues, eq.~\eqref{eqn:HNres}, reproduce \(H_{N-1,\text{EOM}}\).
In terms of the \((t,\{u_i\})\) variables, the poles occur at \(u_a \to 1\) and \(tu_au_b \to 1\) with residues
\begin{eqnarray}
 \left. {\left( {1 - {u_a}} \right)H_N} \right|_{u_a \to 1} &=& \frac{1}{{\prod\limits_{i \ne a} {\left( {1 - {u_i}} \right)\left( {1 - t{u_i}} \right)} }} , \label{eqn:res_uto1} \\
 \left. {\left( {1 - t{u_a}{u_b}} \right)H_N} \right|_{tu_au_b\to 1} &=& \frac{1}{{\left( {1 - {u_a}} \right)\left( {1 - {u_b}} \right)}}\frac{1}{{\prod\limits_{i \ne a,b} {\left( {1 - {u_i}/{u_a}} \right)\left( {1 - {u_i}/{u_b}} \right)} }} ,\label{eqn:res_tuiuj}
\end{eqnarray}
where in the second equation we eliminated \(t\) in favor of \(1/u_au_b\). It is clear that the \(u_a \to 1\) residue coincides with \(H_{N-1,\text{EOM}}\) (see eq.~\eqref{eqn:H_EOM}); to see this in eq.~\eqref{eqn:res_tuiuj}, one needs to rescale the variables \(u_b \to tu_a\) and \(u_i \to t u_i u_a\).

We can understand the above result for the \(u_a \to 1\) limit in terms of the choice of where we put derivatives in the operator basis. Specifically, we can choose to remove all derivatives acting on $\phi_a$ whenever it appears in a term by using IBP and EOM identities; doing so saturates their use so there is no further freedom. In terms of the module, this is associated with an ordering scheme where \(P_1^{(a)} > P_1^{(i)}\) for all \(i\ne a\), \emph{cf.} footnote~\ref{foot:elim_deriv} and its preceding statement. We do not, however, have a simple understanding of the  \(tu_au_b\to 1\) limit in the EFT picture.

It is highly non-trivial that the residue of \(\bou_a\bou_b \to 1\) gives \(H_{N-1,\text{EOM}}\). The polynomial in the numerator of \(H_N\) is quite involved---see eqs.~\eqref{eqn:HN1} and~\eqref{eq:someres2}---and reflects non-trivial relations amongst the generators of the operator basis. In fact, the consistency conditions implied by the pole information completely determines the numerator of $H_N$.\footnote{We thank Bernd Sturmfels and Yeping Zhang for correspondence over our initial conjecture on this point. We are especially grateful to Bernd Sturmfels for proving the conjecture.} Before our understanding of the connection to \(SL(2,\mathbb{C})\), we obtained the residues of \(H_N\) from the sum formula in eq.~\eqref{eqn:HFinal_Sum}. 
Analysis of the residues \(u_a \to 1\) and, in particular, \(tu_au_b\to 1\) allowed us to compute the Hilbert series up to \(N=7\). We note that extracting the residues from the sum in eq.~\eqref{eqn:HFinal_Sum} is manageable, although performing the full sum by brute force is quite difficult for $N>3$.

Starting with the Hilbert series for \(N\) flavors we can obtain the Hilbert series with \(N-k\) flavors by setting \(\bau_{N-k+1}=\dots=\bau_N = 0\). That we can pass to fewer flavors is not too surprising; what's more interesting is that we can also go the opposite direction! In other words, we can compose \(H_N\) from Hilbert series with fewer flavors. This recursion relation can be seen as follows. The $H_{N,\text{EOM}}$ satisfy a trivial recursion relation: $H_{N+1,\text{EOM}}=H_{N,\text{EOM}}\cdot H_{1,\text{EOM}}$. Since $H_{N,\text{EOM}}$ appears in the integrand of eq.~\eqref{eq:molein}, this induces a recursion relation on $H_{N}$. This recursion is easily proved and takes the form
\begin{equation}
{H_{N + 1}}\left( {{\bau_0},{\bau_1}, \cdots ,{\bau_{N + 1}}} \right) = \oint_{\left| x \right| = 1 } {\frac{dx}{{2\pi i}}\frac{{1}}{x}  {H_N}\left( {{\bau_0}, \cdots ,{\bau_{N-1}}} ,x\right)  {H_2}\left( {{x^{ - 1}},{\bau_N},{\bau_{N + 1}}} \right) } . \label{eqn:composition}
\end{equation} 
More generally,  $H_{N+1}\sim\oint \frac{dx}{x} H_{k}H_{k'}$ for any $k,k'$ such that $k+k'=N+2$ and $k,k'\ge2$. 

We can give a graphical description of this composition rule as follows. The basic building block is \(H_2(\bau_0,\bau_1,\bau_2)\), to which we associate a trivalent vertex:
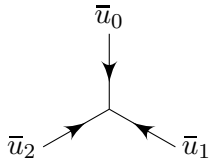
\begin{center}
\begin{tikzpicture}
\begin{scope}[scale=1]
\draw[->-=0.65,arrowhead=2.9mm] (0.87,-0.5)--(0,0);
\draw[->-=0.65,arrowhead=2.9mm] (-0.87,-0.5)--(0,0);
\draw[->-=0.65,arrowhead=2.9mm] (0,1)--(0,0);

\node[anchor=south] at (0,.95) {\(\bou_0\)};
\node[anchor=west] at (0.82,-.5) {\(\bou_1\)};
\node[anchor=east] at (-0.82,-.5) {\(\bou_2\)};
\end{scope}
\end{tikzpicture}
\end{center}
Each leg is associated to a weight \(\bou_i\) with the direction of the arrow indicating whether the weight is taken with a postive power (incoming, \(\bou_i^{+1}\)) or a negative power (outgoing, \(\bou_i^{-1}\)). Higher \(H_N\) are formed by connecting the graphs in such a way that internal lines have the same weight with arrow direction preserved, and then integrating over the weights of the internal lines. For example we can compose two \(H_2\) to get \(H_3\),
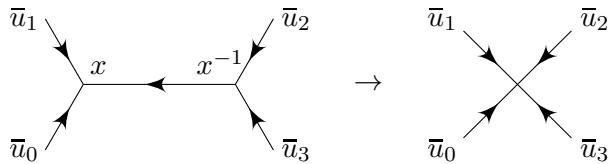
\begin{center}
\begin{tikzpicture}
\draw[->-=0.65,arrowhead=2.9mm] (-.5,0.87)--(0,0);
\draw[->-=0.65,arrowhead=2.9mm] (-.5,-0.87)--(0,0);
\draw[->-=0.6,arrowhead=2.9mm] (2,0)--(0,0);
\draw[->-=0.65,arrowhead=2.9mm] (2.5,0.87)--(2,0);
\draw[->-=0.65,arrowhead=2.9mm] (2.5,-0.87)--(2,0);

\node[anchor=east] at (-.45,0.87)    {\(\bou_1\)};
\node[anchor=east] at (-.45,-0.87)   {\(\bou_0\)};
\node[anchor=west] at (2.47,0.87)    {\(\bou_2\)};
\node[anchor=west] at (2.47,-0.87)   {\(\bou_3\)};

\node[anchor=south] at (.2,0)    {\(x\)};
\node[anchor=south] at (1.8,0)   {\(x^{-1}\)};

\node at (3.75,0) {\(\rightarrow\)};

\begin{scope}[xshift=5.71cm]
\draw[->-=0.65,arrowhead=2.9mm] (-.71,0.71)--(0,0);
\draw[->-=0.65,arrowhead=2.9mm] (-.71,-0.71)--(0,0);
\draw[->-=0.65,arrowhead=2.9mm] (.71,0.71)--(0,0);
\draw[->-=0.65,arrowhead=2.9mm] (.71,-0.71)--(0,0);

\node[anchor=east] at (-.66,0.87)    {\(\bou_1\)};
\node[anchor=east] at (-.66,-0.87)   {\(\bou_0\)};
\node[anchor=west] at (.69,0.87)    {\(\bou_2\)};
\node[anchor=west] at (.69,-0.87)   {\(\bou_3\)};
\end{scope}

\end{tikzpicture}
\end{center}
which reads \(H_3(\bau_0,\bau_1,\bau_2,\bau_3) = \oint \frac{dx}{2\pi i}\frac{1}{x} H_2(\bau_0,\bau_1,x)H_2(x^{-1},\bau_2,\bau_3)\) and defines a new, quartic vertex for \(H_3\). A more elaborate graph is shown in Fig.~\ref{fig:composition}. It is quite clear that all such tree graphs describe the various possible composition formulas for \(H_N\).

\begin{figure}
\centering
\begin{tikzpicture}
\begin{scope}[scale=1.]
\draw[->-=0.6,arrowhead=2.9mm] (1.5,0)--(0,0); 
\draw[->-=0.65,arrowhead=2.9mm] (.31,0.95)--(0,0);
\draw[->-=0.65,arrowhead=2.9mm] (.31,-0.95)--(0,0);
\draw[->-=0.65,arrowhead=2.9mm] (-.81,-0.59)--(0,0);
\draw[->-=0.6,arrowhead=2.9mm] (-1.21,.88)--(0,0); 
\draw[->-=0.65,arrowhead=2.9mm] (-1.12,1.88)--(-1.21,.88);
\draw[->-=0.65,arrowhead=2.9mm] (-2.13,0.47)--(-1.21,.88);

\draw[->-=0.6,arrowhead=2.9mm] (2.25,1.3)--(1.5,0); 
\draw[->-=0.65,arrowhead=2.9mm] (1.38,1.8)--(2.25,1.3);
\draw[->-=0.65,arrowhead=2.9mm] (2.75,2.17)--(2.25,1.3);
\draw[->-=0.6,arrowhead=2.9mm] (3.55,0.55)--(2.25,1.3); 

\draw[->-=0.65,arrowhead=2.9mm] (3.55,-0.45)--(3.55,0.55);

\draw[->-=0.65,arrowhead=2.9mm] (4.42,1.05)--(3.55,0.55);
\draw[->-=0.65,arrowhead=2.9mm] (5.29,0.55)--(4.42,1.05);
\draw[->-=0.65,arrowhead=2.9mm] (4.42,2.05)--(4.42,1.05);

\draw[->-=0.65,arrowhead=2.9mm] (2,-0.87)--(1.5,0);
\end{scope}

\end{tikzpicture}
\caption{Hilbert series can be composed to build up \(H_N\) for larger numbers of flavors.}\label{fig:composition}
\end{figure}


\section{Discussion}
\label{sec:summary}

In this work we have sketched a framework for studying operator bases in quantum field theories and applied this to one-dimensional field theories with scalar degrees of freedom. Our discussion thus far has been fairly mathematical and our one-dimensional application seems rather distant from QFTs of phenomenological and/or theoretical interest. In light of this, it seems prudent to understand what physics lies in our results and what lessons we can extract as we look towards extensions to higher dimensions and more involved QFTs. 

Although we motivated our study of independent operators through the context of effective field theory, our analysis more generally can be understood as classifying the space of local operators subject to some (physically motivated) constraints. Accounting for the equations of motion identifies operators which are equivalent when inserted into correlation functions. The operator-state correspondence suggests a physical meaning to this set for a CFT, although no clear interpretation is immediate for infrared trivial theories. Including integration by parts further restricts to operators of zero momentum, \textit{i.e.} those operators which can contribute to scattering processes. As we review and discuss our results, this picture provides an intuitive understanding for the appearance of mathematical similarities to scattering amplitudes and CFTs---particularly, the role of kinematic equations in our analysis, the representation theoretic description of operators, as well as recursion and composition formulas in the Hilbert series.

In Sec.~\ref{sec:framework}, we showed how the language of commutative algebra provides a systematic way to study operator bases wherein independent operators are understood as elements of a quotient ring. Kinematic equations play an essential role in this framework: momentum conservation and the equations of motion define the equivalence relations governing the quotient space. While each kinematic constraint is separately easy to understand, they have more subtle consequences when considered together. The language of rings and ideals provides a well-defined and systematic way to study the non-trivial relations among these constraints.

The basic details of this framework straightforwardly generalizes to \(d\) dimensions, although explicit computations will differ. Details of this generalization will be discussed elsewhere; here we content ourselves with a few observations. Operationally, we replace \(q\) by \(q_{\m}\) in the ring of momenta and additionally impose invariance of the ring under \(SO(d)\) symmetry. As in one dimension, kinematic equations define an ideal which accounts for redundancies due to IBP and use of EOM. 

For simplicity, let us include parity and instead impose \(O(d)\) symmetry. Invariance under \(O(d)\) implies that the quotient ring consists of polynomials in (symmetric combinations of) the kinematic invariants \(q_i \cdot q_j\), subject to relations stemming from momentum conservation and equations of motion. This is very similar to scattering amplitudes, where Lorentz invariance implies the amplitude can only depend on the invariants \(q_i \cdot q_j\), subject to momentum conservation and on-shell conditions. With what level of seriousness we should take this analogy with amplitudes is unclear to us. In particular, the module contains \textit{polynomials} of the \(q_i \cdot q_j\), while amplitudes are \textit{rational functions} in these invariants whose analytic structure carries deep physical significance. However, there may be non-trivial analytic structure when considering the whole set of polynomials, \textit{i.e.} the entire module. For example, poles and residues lead to intricate consistency conditions in our $d=1$ Hilbert series, as discussed in Sec.~\ref{sec:poles}.

To give an explicit example in \(d\) dimensions, take a single scalar and consider 
operators formed out of \(r\) powers of \(\ph\) fields and an arbitrary number of derivatives. The quotient ring is
\begin{equation}
\bigslant{\mathbb{R}[q_{1\m},\dots,q_{r\m}]^{O(d)\times S_r}}{\left\langle\{q_{1\m}+\dots+q_{r\m}\},q_1^2,\dots,q_r^2\right\rangle}.
\end{equation}
 For \(r =3\) and \(4\), the generators of this module are not too difficult to compute and the result is simple to understand. One finds that, for \(r=3\), the module is trivial: all \(q_i \cdot q_j\) vanish as a consequence of \(q_i^2 = 0\) and \(\sum_{i=1}^3q_{i \m} = 0\). This is directly analogous to the fact that the three-point amplitude for massless particles vanishes on shell. 

For \(r=4\), momentum conservation and EOM reduce the \(q_i\cdot q_j\) to the usual Mandelstam variables \(s\), \(t\), and \(u\), subject to \(s+ t+u =0\). Again, the module reflects structure reminiscent of scattering, this time the familiar kinematics of four-point amplitudes. To complete the study of the \(r=4\) module, we need to impose invariance under \(S_4\) permutations that act on the index \(i\). \(S_4\) permutes the Mandelstam variables according to the defining representation of \(S_3\) lifted to \(S_4\).\footnote{This is a coincidence for \(r=4\). For general \(r\), the \(r(r-1)/2\) kinematic invariants are \(s_{ij} = q_i\cdot q_j\) where \(i = 1,\dots,r\) and \(i \ne j\) since \(q_i^2 =0\) by EOM. Under the symmetric group \(S_r\), the \(s_{ij}\) decompose as \(s = (r) \oplus (r-1,1) \oplus (r-2,2)\) of dimension, 1, \(r-1\), and \(r(r-3)/2\), respectively. We have used the standard notation for labeling irreducible representations of the symmetric group by the partition associated to a specific Young diagram. \((r)\oplus (r-1,1)\) together form the defining, \(r\)-dimensional representation of \(S_r\). Momentum conservation removes this component: dotting \(\sum_{i=1}^rq_{i \m} = 0\) by each \(q_{i \m}\) we get \(r\) Lorentz invariant equations that are permuted under \(S_r\). Therefore, EOM and IBP eliminate all but the components of \(s_{ij}\) transforming in the \((r-2,2)\) representation of \(S_r\).} 
Therefore, the \(S_4\) invariant polynomials are those symmetric in \(s\), \(t\), and \(u\); with the constraint \(s+t+u=0\), these polynomials are freely generated by \(st + su + tu\) and \(stu\). The Hilbert series is \(H_4(t) = 1/(1-t^4)(1-t^6)\) where \(t\) is the weight associated to the derivative (not to be confused with the Mandelstam variable).

In \(d\)-dimensions, for a fixed number of fields we can form an infinite number of operators by application of derivatives. Therefore, unlike the case in one-dimension, we expect the Hilbert series for the fixed number of fields, \(H_r(t)\), to be an infinite series (\textit{e.g.}, \(H_4(t)\) in the above paragraph). Moreover, we anticipate that the full Hilbert series of the EFT will contain an infinite product, reflecting an infinite number of generators in the operator basis. These results are to be anticipated physically as well; by passing to \(d>1\) dimensions we move from quantum mechanics to quantum field theory.

How might the representation theory picture of Sec.~\ref{sec:su2} generalize to more complicated EFTs? For the EFT studied here, once equations of motion are included, the field \(\ph_i\) together with its descendant \(\pd \ph_i\), fill out a representation of \(SL(2,{\mathbb C})\). Once we further impose IBP, we understand the basis to be constructed from the highest weight states in the irreducible decomposition of \(SL(2,{\mathbb C})\) tensor products.  In \(d\) dimensions, however, the number of operators obtained through successive application of \(\pd_{\m}\) to \(\ph_i\) is infinite, even when equations of motion are included---EOM only remove the trace components of the derivatives. This whole picture is reminiscent of primary states in a CFT where descendant states are obtained through application of the lowering operators in the conformal algebra. Our one-dimensional experience and the analogy to CFTs suggests looking for a representation theoretic understanding in \(d\) dimensions as well. 

For this one-dimensional example,  global group structure can be included straightforwardly, using the exposition of the Hilbert series in the form of Molien's formula, eq.~\eqref{eq:molein_notation}. Generalizing this formula when the fields are charged under additional global symmetries proceeds along the lines presented in \textit{e.g.} \cite{Lehman:2015via}. 

The connections between $H_{N,\text{EOM}}$ and $H_{N}$ that we have seen in section~\ref{sec:poles} suggest similar features will persist in more general EFTs and should be looked for; the same can be said for the limits and composition formulas of the Hilbert series we found. For example, the strategy of obtaining $H_{N,\text{EOM}}$  and then searching for a relevant projection to incorporate IBP equivalence may be useful when moving to more complicated EFTs.

To conclude, in this paper we have studied operator bases of EFTs, focussing on one object---the Hilbert series---which encapsulates aspects of the entire operator basis. Requesting a physical basis requires us to take into account EOM and IBP which shape the Hilbert series. The  picture that emerges is that the Hilbert series is an object much akin to the partition function of the theory.
As well as exploring the Hilbert series of more complicated EFTs, it seems worthwhile to search for other objects which can provide information about  operator bases as a whole.


\acknowledgments
We thank Landon Lehman, Adam Martin, Mike Zaletel, and Yeping Zhang for conversations and correspondence, and
Takumi Murayama, Bernd Sturmfels, and Yuji Tachikawa  for valuable comments on an early version
of this manuscript.
TM is supported by U.S. DOE grant DE-AC02-05CH11231 and acknowledges  
computational resources provided through ERC grant number 291377: ``LHCtheory''. 
HM is supported in part by the U.S. DOE under Contract DE-AC03-76SF00098, in part by the
NSF under grant PHY-1316783, in part by the JSPS Grant-in-Aid for
Scientific Research (C) (No.~26400241), Scientific Research on
Innovative Areas (No.~15H05887), and by WPI, MEXT, Japan.


\appendix
\section{A primer on commutative algebra}\label{app:a}

In this appendix we review some basic definitions and results from commutative algebra that we employ in the main text. Useful references include, for example, the introductory text~\cite{Cox:ideals} as well as~\cite{Schenck:2003}, which emphasizes computational aspects through the use of the computer package \texttt{Macaulay2}. In this appendix, \textit{field} takes the traditional mathematical definition, \textit{i.e.} a set that obeys notions of addition and multiplication and their inverses, and has nothing to do with the fields of quantum field theory.

Informally, a \textbf{commutative ring} (herein, ring) is a field without the requirement of a multiplicative inverse. Integers form a ring; rational numbers form a field. Obviously, any field is also a ring. An \textbf{ideal} is a subset of a ring such that the result of multiplying an element of the ideal by an element of the ring remains in the ideal. For example, the even numbers form an ideal of the integers. More formally, let \(R\) be a ring. Then a subset \(I \subset R\) is an ideal if it satisfies i) \(0 \in I\), ii) if \(a,b\in I\) then \(a+b \in I\), and iii) if \(a \in I\) and \(b\in R\) then \(b\cdot a \in I\). 

For our purposes, the most important example of a ring is the \textbf{polynomial ring} \(K[x_1,\dots,x_n]\) consisting of polynomials in the \(x_1,\dots,x_n\) with coefficients in the ring \(K\). In this work \(K\) is typically taken to be a field, such as the real numbers \(\mathbb{R}\). As the addition and multiplication of two polynomials is still a polynomial, quite obviously the polynomials form a ring. A monomial is a term of the form \(x_1^{\a_1}\dots x_n^{\a_n}\); a polynomial is a linear combination of monomials. Intuitively, monomials act like basis elements from which we can build polynomials via addition.

The idea of counting elements in a ring is important to our work, and we anticipate that this somehow reduces to counting monomials. To make this intuition precise, we need the notion of \textit{grading}. 
The polynomial ring \(R = K[x_1,\dots,x_n]\) is naturally graded by degree, where the degree of a monomial is \(\text{deg}(x_1^{\a_1}\dots x_n^{\a_n}) = \a_1 + \dots + \a_n\). A polynomial is \textit{homogeneous} if all of its constituent monomials are of the same degree. Letting \(R_k\) be the set of all homogeneous, degree \(k\) polynomials, then the polynomial ring has a direct sum decomposition \(R = \bigoplus_{k\in \mathbb{N}} R_k\). 
Mathematically, \(x_1,\dots,x_n\) are said to form a \(\mathbb{N}\) graded algebra. 

The dimension of \(R_k\) is simply the number of degree \(k\) monomials. 
For example, in \(R= \mathbb{R}[x,y]\) any homogeneous, degree two polynomial can be written as a linear combination of \(x^2\), \(y^2\), and \(xy\), hence \(\text{dim}(R_2) = 3\). We define the \textbf{Hilbert function} to be \(HF(R,k) = \text{dim}(R_k)\).\footnote{The coefficients \(c_{k\, \mathbf{r}}\) and \(c_k\) defined in the Hilbert series of the main text are Hilbert functions. We avoided this language so as not to over burden those unfamiliar with commutative algebra with terminology.} The \textbf{Hilbert series} of the graded ring \(R\) is then defined as
\begin{equation}
H(R,t) = \sum_k HF(R,k) t^k.
\label{eqn:hs_def_ring}
\end{equation}
For \(R = K[x_1,\dots,x_n]\), the number of degree \(k\) monomials is simply the number of ways of gathering \(k\) items out of \(n\) objects (multiples allowed), \textit{i.e.} 
\[
HF\big(K[x_1,\dots,x_n],k\big) = \sum_{k_1+\dots+k_n=k} = \binom{n+k-1}{k},
\]
and the Hilbert series is
\begin{equation}
H\big(K[x_1,\dots,x_n],k\big) = \sum_{k=0}^{\infty} \binom{n+k-1}{k} t^k = \frac{1}{(1-t)^n}.
\label{eqn:hs_polyring}
\end{equation}
This reflects the fact that there are $n$ generators of this ring, all of degree one, with no relations among them.

Let us now discuss ideals of the polynomial ring. Take \(s\) polynomials in the ring, \(f_1,\dots, f_{s} \in K[x_1,\dots,x_n]\). Then the ideal formed by these polynomials, \(\avg{f_1,\dots,f_{s}}\), heuristically is the set of all polynomials obtained by taking the \(f_i\) as basis vectors where the coefficients \(h_i\) are themselves polynomials in \(K[x_1,\dots,x_n]\). In equations, this set is
\begin{equation}
\avg{f_1,\dots,f_{s}} = \Set{ \sum_{i=1}^{s} h_if_i : \, h_1,\dots,h_{s}\in K[x_1,\dots,x_n] }.
\label{eqn:ideal_def}
\end{equation}
Geometrically, if we imagine that the indeterminates \(x_1,\dots,x_n\) take values in the field \(K\), then the \textbf{variety} \(\mathbf{V}\) defined by the \(f_i\) are the points in \(K^n\) which are solutions to \(f_1 = \dots = f_s = 0\). The ideal \(\avg{f_1,\dots,f_s}\) is then the set of all ``polynomial consequences'' of \(f_1 = \dots = f_s = 0\), \textit{i.e.} the set of all polynomials which vanish on \(\mathbf{V}\). This connection between varieties and ideals is the starting point of the algebra-geometry dictionary.

Let \(R = K[x_1,\dots,x_n]\) be a polynomial ring graded by degree and let \(I = \avg{f_1,\dots,f_s}\) be an ideal of \(R\).  We may quotient the ring by the ideal,
\begin{equation}
M = \bigslant{R}{I}.
\label{eqn:mod}
\end{equation}
By definition, \(M\) consists of equivalence classes of polynomials, where two polynomials are equivalent if they are related by a polynomial in the ideal, \textit{i.e.} for \(h_1,h_2 \in R\) and \(h_3 \in I\), \(h_1 \sim h_2\) if \(h_1 = h_2 + h_3\). In particular, elements of the ideal are equivalent to the zero polynomial and are thus removed from \(M\). The quotient preserves algebraic structure; namely, \(M\) is also a ring. 

In this appendix we will always assume that the polynomials \(f_i\) which define the ideal are homogeneous; in this case, it is clear that the ideal is also graded by degree.\footnote{This is referred to as \textit{projective}, since under a rescaling \(x_i \to \l x_i\) a homogeneous polynomial of degree \(\a\) is simply scaled by \(\l^{\a}\), \(f(x_1,\dots,x_n) \to f(\l x_1,\dots,\l x_n) = \l^{\a}f(x_1,\dots,x_n)\). The term \textit{affine} is used when one or more of the \(f_i\) is not homogeneous. In this work we always are in the projective case, \textit{i.e.} every ideal in the main text is homogeneous in the grading.} In this case, the quotient preserves the grading and \(M\) is said to be a graded module. That is, \(M\) has a direct sum decomposition, \(M = \bigoplus_k M_k\) where \(M_k\) contains the degree \(k\) homogeneous polynomials in \(M\). We can also define a Hilbert function and Hilbert series for the module \(M\). Since \(M_k = R_k/I_k\), the Hilbert function on \(M_k\) is
\begin{equation}
HF(M,k) = \text{dim}(M_k) = \text{dim}(R_k) - \text{dim}(I_k).
\label{eqn:hf_mod}
\end{equation}
The Hilbert series for \(M\) is defined analogously to eq.~\eqref{eqn:hs_def_ring}, \(H(M,t) = \sum_kHF(M,k)t^k\).

To study the quotient ring eq.~\eqref{eqn:mod}, we must first determine properties of the ideal. In a typical situation, we start with some polynomials \(f_1,\dots,f_s\) which generate an ideal. Important questions include determining whether or not a polynomial is in the ideal (ideal membership), possible non-trivial relations among the generators, computing the Hilbert function, \textit{etc.} Answering these inherently computational questions is, in general, difficult.

A simple observation sets us on our way towards computationally probing \(\avg{f_1,\dots,f_s}\). As the \(f_i\) can be thought of as basis vectors for the ideal, it is possible to change bases. In other words, we can find another set of polynomials which generate the same ideal. A particularly nice choice of basis is a \textbf{Gr\"obner basis}, which provides an algorithmically ``best'' way of presenting the polynomial consequences of \(f_1,\dots,f_s\). We denote the set of polynomials in the Gr\"obner basis by \(g_1,\dots,g_r\) (note, \(r \ne s\) in general). By construction, \(\avg{g_1,\dots,g_r} = \avg{f_1,\dots,f_s}\). The algorithm for constructing the Gr\"obner basis is the polynomial generalization of Gaussian elimination familiar from linear algebra.
We make only a few statements pertaining to Gr\"obner bases; a thorough treatment can be found in chapter 2 of~\cite{Cox:ideals}.

To algorithmically manipulate polynomials, an ordering scheme for monomials needs to be chosen. Given two monomials \(\bfx^{\a} = x_1^{\a_1}\dots x_n^{\a_n}\) and \(\bfx^{\b} = x_1^{\b_1}\dots x_n^{\b_n}\), a \textit{monomial order} ``\(>\)'' determines whether \(\bfx^{\a} > \bfx^{\b}\), \(\bfx^{\a} = \bfx^{\b}\), or \(\bfx^{\a} < \bfx^{\b}\).\footnote{There are many possible monomial orders. The three most common are lexographic, graded lexographic, and graded reverse lexographic. For example, graded lexographic order is described as follows. Consider two monomials \(\bfx^{\a} = x_1^{\a_1}\dots x_n^{\a_n}\) and \(\bfx^{\b} = x_1^{\b_1}\dots x_n^{\b_n}\) of total degree \(\a = \a_1 + \dots + \a_n\) and \(\b = \b_1 + \dots + \b_n\), respectively. We consider \(\bfx^{\a} > \bfx^{\b}\) if \(\a > \b\); if \(\a = \b\), then \(\bfx^{\a} > \bfx^{\b}\) if \(\a_1 > \b_1\); if \(\a = \b\) \textit{and} \(\a_1 = \b_1\), then  \(\bfx^{\a} > \bfx^{\b}\) if \(\a_2 > \b_2\); and so on.} Moreover, a given monomial order allows us to specify the ``largest'' term for a polynomial \(h \in K[x_1,\dots,x_n]\), which we call the initial monomial of \(h\) and denote by \(\text{in}(h)\).\footnote{This is also commonly called the leading term of \(h\) and denoted by \(\text{LT}(h)\).} For an ideal \(I = \avg{f_1,\dots,f_s}\), we take the initial monomials of every polynomial in \(I\) and denote this set by \(\text{in}(I)\). Note that, in general, \(\text{in}(I)\) is \textit{not} equal to the set generated by the initial monomials of the \(f_i\). In fact, \textit{the defining property of a Gr\"obner basis} is that \(\text{in}(I) = \avg{\text{in}(g_1), \dots, \text{in}(g_r)}\).

The Hilbert series of quotient rings is one of the simpler objects one can compute---as it counts independent monomials, it does not require knowing full information about the elements in a module. In regards to computing the Hilbert series, one of the nice properties of Gr\"obner bases is that we can compute the initial monomials of the ideal, \(\text{in}(I)\), from the initial monomials of the Gr\"obner basis.

Let us now give a few explicit examples to highlight some of the concepts introduced in this appendix. We take the polynomial ring in two variables with coefficients in the real numbers, \(R = \mathbb{R}[x,y]\), and consider ideals that are similar to those of the main text. Various computer packages can be used to calculate the Gr\"obner basis; the next appendix gives an example using \texttt{Macaulay2}.\footnote{\texttt{Mathematica} computes a Gr\"obner basis via the command \texttt{GroebnerBasis[\{polynomials\},\{variables\}]}.}  For the monomial order, we use graded reverse lexographic ordering, which is the default for the computer package \texttt{Macaulay2}. 

\noindent \textbf{Example 1}\\
\noindent Let \(R = \mathbb{R}[x,y]\) and \(I = \avg{x+y}\). As the ideal consists of a single polynomial, it is already a Gr\"obner basis. Hence, the initial ideal is generated by the \(\text{in}(x+y) = x\), \(\text{in}(I) =\avg{x}\). The Hilbert series of the quotient ring \(M = R/I\) is equivalent to the Hilbert series of \(R/\text{in}(I) = \mathbb{R}[x,y]/\avg{x} = \mathbb{R}[y]\). Hence, 
\[
H(\mathbb{R}[x,y]/\avg{x+y},t) = \frac{1}{1-t}\, .
\]

\noindent \textbf{Example 2}\\
\noindent Let \(R = \mathbb{R}[x,y]\) and \(I = \avg{x^2,y^3}\). A monomial \(x^{\a}y^{\b}\) is quite clearly in the ideal for \(\a \ge 2\) or \(\b \ge 3\). Then the monomials of the quotient ring are \(1,x,y,xy,y^2,\) and \(xy^2\). The Hilbert series is then
\begin{equation*}
 H(\mathbb{R}[x,y]/\avg{x^2,y^3},t) = 1 + 2t + 2t^2 + t^3 \, .
\end{equation*}
It is perhaps illuminating to recognize that
\[
H = \frac{(1-t^2)(1-t^3)}{(1-t)^2} = \frac{1 -t^2 - t^3 +t^5}{(1-t)^2},
\]
where \(1/(1-t)^2\) is the Hilbert series of the free ring \(\mathbb{R}[x,y]\), while the numerator reflects information about the generators of the ideal and the relations among them.

\noindent \textbf{Example 3}\\
\noindent Let \(R = \mathbb{R}[x,y]\) and \(I = \avg{x+y,x^2,y^3}\). The Gr\"obner basis is given by \(x+y\) and \(y^2\), \(\avg{x+y,x^2,y^3} = \avg{x+y,y^2}\). Restricting to the initial ideal, \(\text{in}(I) = \avg{x,y^2}\). In the quotient ring, the basis monomials are \(1\) and \(y\) and the Hilbert series is
\begin{equation*}
 H(\mathbb{R}[x,y]/\avg{x+y,x^2,y^3},t) = 1 + t .
\end{equation*}


\section{Macaulay2 demo: enumerating and constructing operator bases}
\label{app:b}
This is a demonstration of how calculate the Hilbert series (which enumerates the independent operators), and an explicit realization of an independent set of operators, using the program {\tt Macaulay2}~\cite{m2}. 

We consider the $(0+1)$ dimensional EFT of three flavors ($N=3$) of real scalar fields $\phi_i$, and show how to enumerate and find an independent set of operators of the form $\phi_1^3\, \phi_2^4\,\phi_3^5 \,\dd^k$; we consider the most general case by allowing for any number, $k$, of derivatives in our counting.

Following section~\ref{sec:framework}, we wish to compute the Hilbert series of the module ${\mathbb R}[x,y,z]/\langle x+y+z,x^4,y^5,z^6\rangle$.
This Hilbert series is obtained in {\tt Macaulay2} via the following commands:\newline
{\tt i1:    R=QQ[x,y,z]; }  \newline
{\tt i2:    I=ideal(x+y+z,x\^{}4,y\^{}5,z\^{}6); }\newline
{\tt i3:    hilbertSeries(R/I, Reduce=>true) }\newline
In the first line, {\tt QQ} means that the ring is taken over the field of rational numbers. The {\tt Reduce=>true} option factors the Hilbert series. The output obtained from the above is $1 + 2t + 3t^2  + 4t^3  + 4t^4  + 3t^5 + t^6$. The power of $t$ in the output counts the number of derivatives in the operator: one independent operator with no derivatives, two independent operators with one derivative, three independent operators with two derivatives, {\it etc.}. No operators survive with more than six derivatives---the EOM render the series finite. 

We note in passing that a Gr\"obner basis for this ideal can be output with:\newline
{\tt i4:    gens gb I }\newline
{\tt o4 = x+y+z y4+4y3z+6y2z2+4yz3+z4 10y3z2+20y2z3+15yz4+4z5 z6 5y2z4+6yz5 }\newline
where in the last line, {\tt 5y2z4+6yz5} is read as $5 y^2z^4+6yz^5$ {\it etc.}. There are five polynomials in the Gr\"obner basis.

To further construct an explicit basis, we proceed via the following commands:\newline
{\tt i5:    T=R/I; }\newline
{\tt i6:    sort basis T }\newline
{\tt o6= 1 z y z2 yz y2 z3 yz2 y2z y3 z4 yz3 y2z2 y3z z5 yz4 y2z3 yz5 }\newline
where the last line is output. This output, translated back to the corresponding operators, provides the set of independent operators (sorted by the number of derivatives they contain): $\phi_1^3\, \phi_2^4\,\phi_3^5$, $~~\phi_1^3\, \phi_2^4\,\phi_3^4 (\dd \phi_3)$, $~~\phi_1^3\, \phi_2^3(\dd\phi_2) \,\phi_3^5$,  $~~\phi_1^3\, \phi_2^4 \,\phi_3^3 (\dd \phi_3)^2$,  $~~\phi_1^3\, \phi_2^3 (\dd\phi_2) \,\phi_3^4 (\dd \phi_3)$,  $~~\phi_1^3\, \phi_2^2(\dd\phi_2)^2 \,\phi_3^5$, {\it etc.}.

\bibliography{./bibliography}
\bibliographystyle{JHEP}

\end{document}